\RequirePackage{fix-cm}
\documentclass[natbib,twocolumn]{svjour3}          
\smartqed  
\usepackage{graphicx}
\usepackage[bookmarks = true, bookmarksnumbered = true, pdfpagemode =None, pdfstartview = FitH, pdfpagelayout = SinglePage, colorlinks = true, urlcolor = blue, citecolor = blue]{hyperref}
\usepackage{lineno}

 \journalname{Space Science Reviews}
\begin{document}

\title{How well do we understand the belt/zone circulation of Giant Planet atmospheres?
}
\subtitle{}

\titlerunning{Belts and Zones}        

\author{Leigh N. Fletcher         \and
        Yohai Kaspi \and
        Tristan Guillot \and 
        Adam P. Showman 
}


\institute{L.N. Fletcher \at
              Department of Physics and Astronomy, University of Leicester, University Road, Leicester, LE1 7RH, United Kingdom \\
              Tel.: +44-116-252-3585\\
              \email{leigh.fletcher@le.ac.uk}           
           \and
           Y. Kaspi \at
            Department of Earth and Planetary Sciences, Weizmann Institute of Science, Rehovot 76100, Israel           
           \and
           T. Guillot \at
            Universit\'{e} C\^{o}te d'Azur, OCA, Lagrange CNRS, 06304 Nice, France
           \and
           A.P. Showman \at
            Lunar and Planetary Laboratory, University of Arizona, Tucson, AZ 85721-0092, USA.          
            }

\date{Received: 2019-05-08 / Accepted: 2019-12-24}

\maketitle

\tableofcontents


\begin{abstract}
The atmospheres of the four giant planets of our Solar System share a common and well-observed characteristic:  they each display patterns of planetary banding, with regions of different temperatures, composition, aerosol properties and dynamics separated by strong meridional and vertical gradients in the zonal (i.e., east-west) winds.  Remote sensing observations, from both visiting spacecraft and Earth-based astronomical facilities, have revealed the significant variation in environmental conditions from one band to the next.  On Jupiter, the reflective white bands of low temperatures, elevated aerosol opacities, and enhancements of quasi-conserved chemical tracers are referred to as `zones.'  Conversely, the darker bands of warmer temperatures, depleted aerosols, and reductions of chemical tracers are known as `belts.' On Saturn, we define cyclonic belts and anticyclonic zones via their temperature and wind characteristics, although their relation to Saturn's albedo is not as clear as on Jupiter.  On distant Uranus and Neptune, the exact relationships between the banded albedo contrasts and the environmental properties is a topic of active study.  This review is an attempt to reconcile the observed properties of belts and zones with (i) the meridional overturning inferred from the convergence of eddy angular momentum into the eastward zonal jets at the cloud level on Jupiter and Saturn and the prevalence of moist convective activity in belts; and (ii) the opposing meridional motions inferred from the upper tropospheric temperature structure, which implies decay and dissipation of the zonal jets with altitude above the clouds.  These two scenarios suggest meridional circulations in opposing directions, the former suggesting upwelling in belts, the latter suggesting upwelling in zones.  Numerical simulations successfully reproduce the former, whereas there is a wealth of observational evidence in support of the latter.  This presents an unresolved paradox for our current understanding of the banded structure of giant planet atmospheres, that could be addressed via a multi-tiered vertical structure of ``stacked circulation cells,'' with a natural transition from zonal jet pumping to dissipation as we move from the convectively-unstable mid-troposphere into the stably-stratified upper troposphere.

\end{abstract}

\section{Introduction}
\label{intro}

How well do we really understand the atmospheres of the four giant planets of our Solar System?  As astronomers have been rapidly expanding the census of large, gaseous objects throughout our galaxy over the past two decades (both exoplanets and Brown Dwarfs), the atmospheres of the gas and ice giants in our own solar neighbourhood should serve as archetypes for those that we can never hope to view up close.  But despite decades of ambitious interplanetary spacecraft, combined with long records of Earth-based astronomical observations, there remain a number of fundamental questions and ambiguities about the dynamical and chemical processes shaping the atmospheres of the four giants.  This review will focus on one specific aspect of their circulation - reconciling the banded patterns of clouds, temperature, and composition that characterise all four of these worlds with the mechanisms responsible for both maintaining and impeding the zonal (east-west) jet systems.  We aim to write an observationally-driven review that explains, in general terms, the planetary banding that is observed in multi-wavelength remote sensing.  Given that our knowledge of the gas giants (via Galileo, Juno, and Cassini) significantly exceeds that of the ice giants, we have tried to express caution wherever appropriate.  We also avoid a comprehensive review of giant planet atmospheric dynamics, for which the reader is referred to the discussions in \citet{04ingersoll}, \citet{05vasavada}, \citet{09delgenio}, \citet{18sanchez_jets} and \citet{18showman}.  Instead, this review aims to bring the potential paradox into sharper focus, so that it might be better addressed in future by both numerical simulations \textit{and} planetary observers.

Planetary banding is certainly not unique to the four giant planets.  Earth's general circulation can be subdivided into three distinct latitudinal regimes:  the tropical Hadley circulation characterised by moist convective activity over small scales; the extra-tropical Ferrel circulation characterised by large-scale baroclinic waves; and the polar circulation cell.  Air in the tropics rises due to convergence of a \textit{thermally-direct} circulation, where warm air rises and cold air sinks \citep[Section 10 of][]{04holton}.  The rising air deflects eastward due to the attempt to conserve angular momentum as it moves poleward, thus developing an eastward velocity to form the sub-tropical jet stream at 12-15 km altitude.  The return flow at the surface is deflected towards the west due to the Coriolis force to form the Trade Winds.  In the extra-tropics, the generation of eddies on $\sim1000$-km scales provides a `stirring' mechanism to generate larger-scale Rossby waves \citep{06vallis}.  Rossby waves have the special property that as they propagate latitudinally away from their generation region, they transport momentum back into their formation region (i.e., breaking Rossby waves deposit prograde angular momentum in their source regions, and retrograde angular momentum in their dissipation regions).  This convergence of angular momentum drives the formation of an extra-tropical eastward jet, and this Ferrel-cell circulation is an example of a \textit{thermally-indirect} flow, driven by geostrophic turbulence in the atmosphere\footnote{Although the Ferrel cell circulation is thermally indirect, it is overwhelmed by the heat transport associated with eddies, meaning that the total circulation in this latitude band remains thermally direct, transporting heat towards the poles.}.  Thus Earth's circulation can be subdivided into tropical and extra-tropical regimes, although the boundary between these two is often complex and blurred, as there is not enough `dynamical room' for both the sub-tropical jet and the mid-latitude eddy-driven jet between Earth's equator and pole.  In simple terms, the meridional extent of the Hadley cell depends upon the planetary rotation rate \citep{80held, 15kaspi}, with slow rotators like Titan having global-scale `tropics' and fast-rotators like Jupiter have narrow tropics and multiple Ferrel-like extratropical cells.  This review will deal with observational characteristics of the meridional (i.e., latitudinal) circulations associated with the tropical (potentially Hadley-like) and extra-tropical (potentially Ferrel-like) circulations of the giant planets.

This observational review is organised as follows.  In Section \ref{obs} we describe the key observational constraints (temperatures, gaseous distributions, zonal winds, lightning) that must be explained by any reasonable model of giant planet circulation.  Section \ref{topcell} discusses evidence for the decay of zonal winds with increasing altitude, and the implications and evidence for upwelling in zones and subsidence in belts.  This is followed in Section \ref{lowercell} by the counter-argument, exploring the implications of eddy-driven jets and the distribution of lightning in the extra-tropical regions.  Possible mechanisms to reconcile these different perspectives are discussed in Section \ref{two-cell}, alongside the sparse observational evidence for and against such a complex meridional circulation pattern.  Finally, Section \ref{large-scale} describes how these ideas of multi-tiered cells might be applied to the ice giants, Uranus and Neptune.  

\section{Observations of Belt/Zone Contrasts}
\label{obs}

\begin{figure*}
\begin{centering}
\centerline{\includegraphics[angle=0,scale=.40]{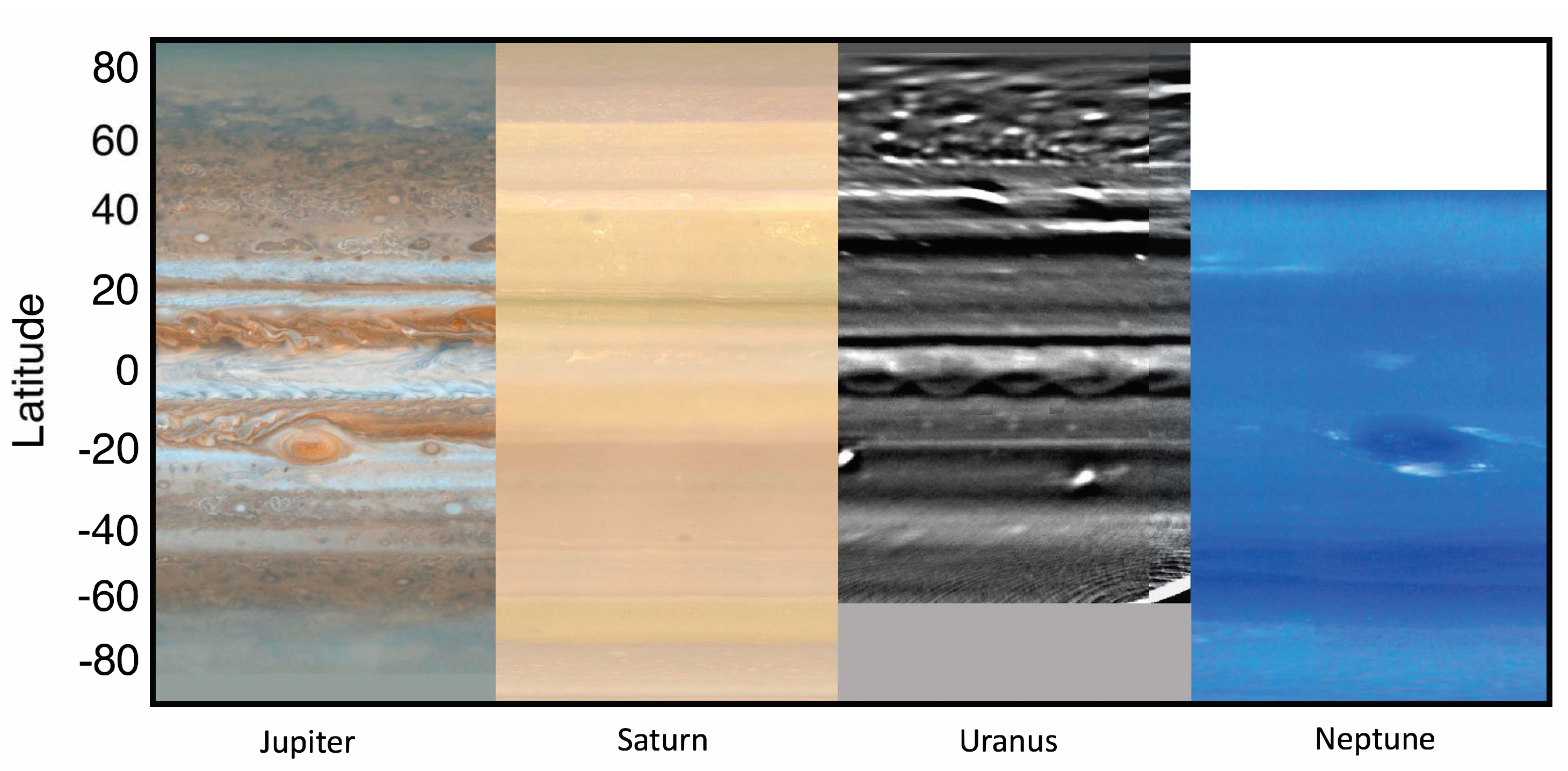}}
\caption{A selection of cylindrically-mapped images to showcase the shared planetary banding on the four giant planets.  The Jupiter map comes from Cassini/ISS images (NASA image PIA07782); the Saturn and Neptune maps were prepared from Cassini and Voyager-2 imaging by Bj\"{o}rn J\'{o}nsson; the Uranus H-band image was obtained by Keck in November 2012 \citep{15sromovsky}. }
\label{images}
\end{centering}
\end{figure*}

The striped appearances of the giant planets, particularly Jupiter, are the dominant characteristics of these worlds (Fig. \ref{images}).  The first observations of zonal banding on Jupiter may have come as early as 1630, a mere 20 years after the invention of the telescope, by Niccolo Zucchi, a Jesuit theologian \citep{98hockey, 95rogers}.  On Saturn,  J.D. Cassini first reported an equatorial belt in 1676\footnote{\url{https://doi.org/10.1098/rstl.1676.0033}}, \textit{``we have discerned on the globe of Saturn a dusky zone [zona obscura], a little farther south than the centre, similar to the zones of Jupiter"} \citep{62alexander}.  Observations of the banded structure of the ice giants had to await the spacecraft age (primarily Voyager), but were later supplemented by observations from powerful ground-based and space-based observatories (e.g., Fig. \ref{images}).  

For historical reasons, the high-albedo bands are referred to as `zones' and the low-albedo regions as `belts,' although these distinctions are currently more meaningful for the Jupiter than any of the other giants.  Indeed, in this review we define belts and zones via their vorticity (e.g., their zonal winds) and their temperatures, rather than by their colours and reflectivity, although the two are often intertwined.  Furthermore, although Jupiter's colour contrasts are observed in the topmost clouds in the 400-800 mbar range (condensates of ammonia, mixed with unidentified chromophores), the belt/zone contrasts extend both upwards and downwards.  Above the condensate clouds, the \textit{upper troposphere} is the stably-stratified region above the radiative-convective boundary (around 300-500 mbar on Jupiter and Saturn), where UV-induced photochemistry can produce aerosols (`hazes') and new chemical products, and the statically-stable temperature profile is controlled by the balance between absorption of sunlight on these aerosols and methane gas, and the emission from hydrocarbons in the stratosphere (Section \ref{topcell}).  Conversely, the temperature profile in the cloud-forming region, or `weather layer,' follows a moist adiabat due to the release of latent heat as the clouds form.  We consider this \textit{mid-troposphere} to extend to the base of the water clouds and to be the location of the primary meteorological features that are likely to be the source of the eddies providing momentum to the zonal jet system (Section \ref{lowercell}).  Beneath the weather layer, the \textit{lower troposphere} is expected to be a much deeper dry-convective layer, with temperatures following the dry adiabat.  Jets and zonal banding do persist into these abyssal layers \citep{13kaspi, 18kaspi, 18guillot, 19galanti, 19iess}, and the lower layer can influence the weather layer by injecting moist-convective elements from the dry-convective layer \citep{16thomson, 07showman, 10lian}, where it must pass through a region of stable stratification (broader beneath the zones than the belts) at the base of the water cloud \citep{14sugiyama}. 

\subsection{Winds and Temperatures}

\begin{figure*}
\begin{centering}
\centerline{\includegraphics[angle=0,scale=.70]{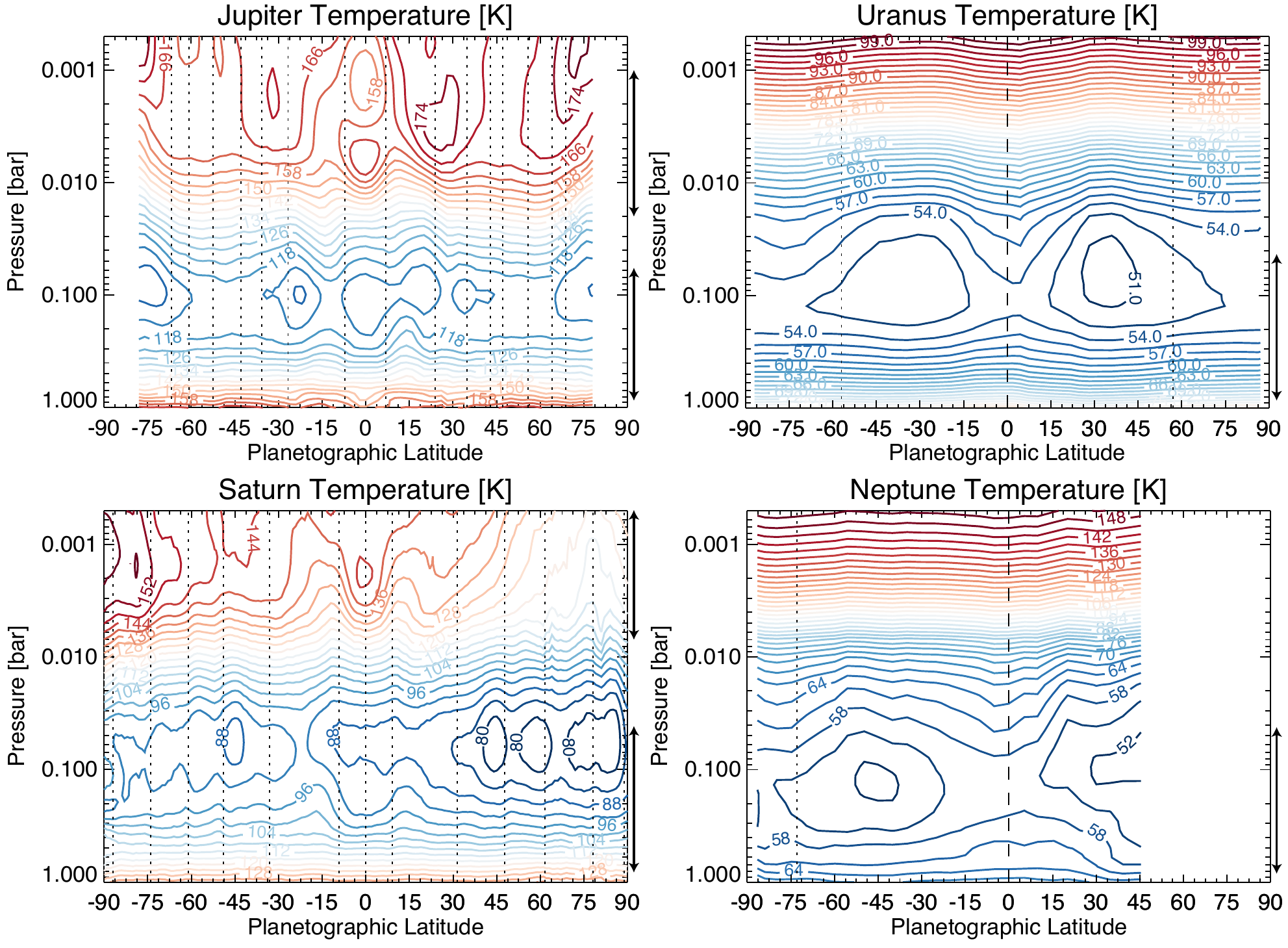}}
\caption{Upper tropospheric and stratospheric temperatures retrieved from thermal infrared observations from Cassini/CIRS (Jupiter and Saturn) and Voyager/IRIS (Uranus and Neptune) using the NEMESIS optimal estimation retrieval algorithm \citep{08irwin}.  Jupiter's temperatures were measured during the December 2000 flyby \citep{09fletcher_ph3}; Saturn's temperatures are averaged from 2006 to 2010, either side of the northern spring equinox \citep{17fletcher_QPO}; Uranus' temperatures were measured during the 1986 flyby (near southern summer solstice) and lack any constraint in the stratosphere \citep{15orton}; Neptune's temperatures were measured in 1989 \citep{14fletcher_nep} but lack the warm south polar vortex that emerged nearer southern summer solstice in 2005.  Temperatures are in geostrophic balance with the zonal winds, with prograde jets indicated by vertical dotted lines.  Uranus' and Neptune's equatorial retrograde jet shown by a dashed line.  Arrows to the right of each plot show the approximate altitude sensitivity of the CIRS and IRIS spectra - domains outside of this range are from a smooth relaxation to an \textit{a priori} profile. Zonal wind peaks are taken from \citet{03porco}, \citet{11garcia} and \citet{18sanchez_jets}.}
\label{temp}
\end{centering}
\end{figure*}

\textit{At the cloud tops: }  Returning to the observable cloud-tops, the meridional (north-south) albedo patterns on Jupiter and Saturn are related - but not always obviously - to their zonal (east-west) wind fields.  Eastward (prograde) jets are observed on the equatorward side of belts, and westward (retrograde) jets on their poleward sides (Fig. \ref{temp}), with potential vorticity gradients associated with the eastward jets themselves serving as barriers to mixing between the different bands \citep[e.g.,][]{06read_jup, 09read}.  Defined in this way, using vorticity rather than the cloud reflectivity, zones possess anticyclonic vorticity, and belts are cyclonic.  Each hemisphere of Jupiter has 6-7 eastward (prograde) jets separated by westward (retrograde) jets, whereas each hemisphere of Saturn has 4-5 eastward jets (Fig. \ref{temp}).  Jupiter's zonal winds have been studied extensively from ground- and space-based facilities since the time of the Voyager missions \citep[e.g., see the extensive review of giant planet winds by][]{18sanchez_jets}, revealing sharp jets of near equal magnitude in both the eastward and westward directions. Conversely, Saturn's eastward jets tend to be sharper than the more rounded zonal-wind minima (the weak westward jets), and strongly biased in the eastward direction in the System III longitude system \citep[e.g.,][]{05porco, 06vasavada, 11garcia}.  However, we note that the System III rotation rate determined by Voyager may not accurately represent the rotation of Saturn's interior - a faster rotation rate has been suggested by several independent arguments \citep{07anderson, 09read_rot, 15helled, 19mankovich}, which would reduce Saturn's bias towards prograde jets and make the velocity profile appear more `Jupiter-like.'

If we retain the Voyager-derived System III, then Saturn's equatorial jet is broader and stronger than Jupiter's.  Both planets exhibit a similar double-peaked wind structure about the equator, with off-equatorial maxima in the eastward winds.  Saturn's equatorial jet also exhibits a narrow-peaked jet that increases in velocity with altitude located right at the equator \citep{11garcia}.  There is no specific reason to expect a complete symmetry of the jets of Jupiter, and indeed we find that Jupiter's zonal winds are asymmetric, with (i) a strong eastward jet at 24$^\circ$N that has no southern counterpart; (ii) the presence of the Great Red Spot in the southern hemisphere, with no northern counterpart; and (iii) an apparent asymmetry in the northern and southern equatorial jets, resulting from the presence of a trapped Rossby wave on the jet between the equatorial zone and the North Equatorial Belt \citep[NEB, e.g.,][]{90allison, 11asaydavis}.  

\textit{Below the clouds: } Studies of the giant planet winds have primarily used the tracking of cloud features in this top-most cloud deck in the 0.5-1.0 bar region.  Observations below these clouds are challenging.  Jupiter's winds were measured by the Galileo probe in a single location \citep{98atkinson}, and were found to increase with depth in Jupiter's tropics down to $\sim4$ bar, and then to remain approximately constant (or show a very weak decrease) to 21 bar, where the probe signal was lost.  On Saturn, observations at 5-$\mu$m by Cassini/VIMS could probe down to the 2-3 bar level, and revealed that tropical winds were generally stronger at depth than at the 500-mbar level, whereas extra-tropical winds were slightly weaker at depth \citep{09baines_pole, 09choi, 11garcia, 11li, 18studwell}, although \citet{18sanchez_jets} cautions that this extra-tropical trend is relatively weak and may not be universal across all of Saturn's jets. Nevertheless, this contrast between the tropics and extra-tropics is important, as we shall see later, and suggests that the winds should decay away with increasing depth.  This decay with depth must be relatively gentle, as recent results have shown that the zonal jets on Jupiter and Saturn penetrate to depths of 3000 km \citep{18kaspi} and 9000 km \citep{19galanti, 19iess}, respectively; and that the depths of the winds on the ice giants can be no more than 1000 km \citep{13kaspi}.  Moreover, the Juno gravity results imply that the same zonal wind pattern extends to great depths \citep{18kaspi}.  Below these depths, it is suggested that the potential for ohmic dissipation of induced currents (i.e., magnetohydrodynamic drag) prevents penetration into the more electrically conductive fluid layers \citep{08liu, 17cao}, meaning that the deeper layers do not experience differential rotation \citep{18guillot}.

\textit{Above the clouds: } Perhaps surprisingly, our knowledge of the zonal winds above the clouds is almost as uncertain as our understanding of the flows at depth.  Given the absence of available cloud tracers in the upper troposphere, the winds must be inferred indirectly from measurements of temperatures.  The thermal wind equation \citep[for an atmosphere in geostrophic balance between the Coriolis acceleration and pressure gradients,][]{04holton} dynamically links winds to gradients in the temperatures and densities.  Observations of Jupiter over four decades \citep[e.g.,][]{83conrath} have shown the cyclonic belts to be warm and depleted in aerosols and gases, suggestive of high-entropy `dry' air sinking downwards.  Anticyclonic zones are colder and display enhancements in aerosols and gases, suggestive of `moist' air rising upwards.  The maximum temperature gradient is therefore located at the belt/zone boundary (Fig. \ref{temp}), causing vertical shear on the jets via the thermal wind equation \citep{04holton}.  This is equally true on Jupiter \citep[e.g.,][]{06simon} and Saturn \citep[e.g.,][]{07fletcher_temp}, but is less clear for the ice giants.  The sense of the windshear is such that the zonal jets will weaken with altitude through the upper troposphere, but in some cases can flip direction and strengthen into the stratosphere \citep[e.g.,][]{09read}.  This windshear, acting against the direction of the cloud-top jets, will be crucial in Section \ref{topcell}.  

\subsection{Clouds and Hazes}

Although the correspondence between the zonal jets and the tropospheric temperature banding is clear for both Jupiter and Saturn (Fig. \ref{temp}), their connection to the aerosol distributions are less obvious.  Albedo contrasts in Fig. \ref{images} are most likely related to changes in the altitude, thickness, and optical properties of the aerosols in (i) the condensate clouds formed from key volatiles (NH$_3$, H$_2$S, and H$_2$O, in addition CH$_4$ on the cold ice giants) and (ii) photochemically-produced hazes from tropospheric (e.g., PH$_3$, NH$_3$) and stratospheric (hydrocarbons like ethane and acetylene) chemistry. On Jupiter there is a close correspondence between red-brown visible colours in the belts and an absence of clouds sensed in the 1-4 bar range \citep[sensed at 5 $\mu$m,][]{04ingersoll,17fletcher_seb,19antunano}, as well as an absence of upper-tropospheric aerosols near 500 mbar \citep[sensed at 8-9 $\mu$m,][]{09fletcher_imaging}.  Conversely, Jupiter's visibly-white zones appear dark at thermal infrared wavelengths, suggesting thick and high clouds over these zones.  Saturn's albedo contrasts are more subdued, primarily due to the thickness of the overlying tropospheric haze - there is a maximum in upper-tropospheric aerosol opacity at the equator \citep[$\pm10^\circ$ latitude,][]{11fletcher_vims, 13roman} that is co-located with a minimum in the temperatures \citep{07fletcher_temp}, but uncertainties on aerosol opacity measurements dwarf the belt/zone contrasts at higher latitudes.  Nevertheless, the clouds and chemical species certainly display a banded structure, even if the correspondence with the temperatures and winds remains unclear.  

In particular, Saturn displays a series of fine latitudinal bands in both the visible \citep{06vasavada} and at 5-$\mu$m \citep{06baines, 09choi}, quite unlike that seen at Jupiter. Eastward jets coincide with dark regions in continuum-band wavelengths (see Fig. \ref{images}), separated by broader and brighter bands associated with the westward jets.  This is also unlike the thermal field \citep{07fletcher_temp}, which displays the same correspondence between winds and temperatures as those seen on Jupiter.   The brightness of Saturn's bands can change with time \citep{06perez-hoyos} and with wavelength, perhaps hinting at complex vertical and temporal variability in Saturn's banded structure.  \citet{09delgenio} point out that Saturn's tropospheric temperature gradients are morphologically similar to the 890-nm albedo patterns, sensing the upper tropospheric hazes.  This hints at a change in the character of the circulation from Saturn's upper troposphere to the cloud decks \citep{09delgenio}, something which we shall explore below.

The correspondence between winds, temperatures, and clouds is even more unclear for the ice giants.  Uranus and Neptune do exhibit zonally-organised albedo patterns \citep[e.g., see the comprehensive review by][]{18sanchez_jets}.  However, these appear to be disconnected from the large-scale wind field, with their equatorial retrograde jets and polar prograde jets \citep{91limaye, 93sromovsky, 98karkoschka, 09sromovsky, 11karkoschka, 15sromovsky}.  They are also disconnected from the thermal field (Fig. \ref{temp}), where mid-latitude upwelling produces cooler temperatures, contrasted with subsidence at the warmer equator and poles \citep{87flasar, 98conrath, 14fletcher_nep, 15orton}.  The small-scale latitudinal banding on Uranus \citep[e.g., Fig. \ref{images} and ][]{12fry, 15sromovsky} appears to have no direct connection to the measured circulation systems from the temperatures and winds, and little is known about any eddy components or meridional motions \citep[see the detailed review by][]{18sanchez_jets}.  Could this imply that the albedo bands are being observed at a different altitude compared to the winds and temperatures?  Or that the spatial resolution of our wind/temperature measurements are simply insufficient to detect contrasts on the scale of the Uranian stripes?  We shall return to these questions in Section \ref{large-scale}.


\subsection{Gaseous Distributions}

\textit{Ammonia cycle: } Perhaps the most important evidence for spatial variations in vertical mixing comes from the distributions of gaseous tracers, both volatiles and disequilibrium species.  Their vertical distributions are governed by the strength and direction of vertical motions, and by the locations of sources (e.g., chemical production, evaporating precipitates) and sinks (condensation, photochemical destruction).  Jupiter and Saturn have two primary condensable cycles, based on ammonia (NH$_3$) and water (methane also condenses on the cold ice giants).  Ammonia is greatly enhanced at the equator on both Jupiter \citep{06achterberg, 16depater, 16fletcher_texes, 17li} and Saturn  \citep{11fletcher_vims, 13janssen, 13laraia}, and depleted in the neighbouring belts.  These enhancements could potentially be connected to the observed distributions of discrete equatorial plumes of ammonia and their associated downdrafts on Jupiter \citep{16fletcher_texes, 16depater} and Saturn \citep{13janssen}, rather than continuous uplift of ammonia at all longitudes.  Ammonia contrasts associated with the belts and zones are most evident in the tropics, and are much more subdued at extra-tropical latitudes (e.g., at the limit of our ability to derive from remote sensing observations).  

Contrary to the expectations of equilibrium condensation schemes, ammonia isn't simply well-mixed up to its cloud base (0.5-1.0 bar for Jupiter and Saturn).  Earth-based observations showed that ammonia is generally depleted down to at least the 6-bar level \citep{01depater}, and Juno revealed that this depletion extends down to at least 40-60 bar \citep{17li}.  Jupiter's water cloud base is expected to be somewhere in the 4-9 bar range, but the equatorial ammonia enhancement extends to at least 40-60 bars \citep{17ingersoll}.  They point out that ammonia-rich air rising, and ammonia-poor air falling, cannot be in steady state - some process needs to close this circulation as rain does for water uplift at Earth's low latitudes.  Ammonia snow would fall through the NH$_3$ cloud base into the warmer air below, eventually sublimating somewhere around 4 bar, re-releasing ammonia into the gas phase \citep{19li}. To get the NH$_3$ all the way back down to 40-60 bar would require ``hidden and dense'' ammonia vapour downdrafts within the equatorial zone \citep{17ingersoll}.  In fact, downdrafts associated with the plumes would be dry and volatile-depleted, the net effect being to deplete the mid-troposphere of ammonia down to the 10-bar level \citep{19li}.  It therefore remains challenging to deplete ammonia below the water cloud (10 bars), so the model of \citet{19li} still cannot explain the apparent depletion down to 40-60 bars observed by Juno \citet{17ingersoll}.

The horizontal NH$_3$ distribution, with the equatorial ammonia maximum on both Jupiter and Saturn, is just as complex as the vertical distribution.  Observations from the microwave and radio \citep{17li, 19depater_vla} show the Northern Equatorial Belt (NEB) to be significantly more depleted than the Southern Equatorial Belt (SEB), an asymmetry potentially driven by the presence of the Great Red Spot in the south, but not in the north.  Furthermore, the SEB is often observed to be broken into narrower albedo bands - NH$_3$ is mostly depleted in the northern and southern components; but moderately enhanced over the core of the SEB - similar latitudinal structure within the SEB has been observed during times when the belt whitened or `faded' \citep{11fletcher_fade}, hinting at a shift in the meridional circulation of the SEB and cooling (and condensation) in the centre of the belt.  The NEB, conversely, has a trend for more NH$_3$ towards its northern edge, and the strongest depletion at the southern edge, near to the hotspots that are a manifestation of the equatorially-trapped Rossby wave.  
    
\textit{Disequilibrium species: }  In addition to ammonia, a number of gaseous species can be considered to be tracers of upwelling and subsidence.  In the stratosphere, hydrocarbons trace larger-scale circulation patterns and inter-hemispheric transport \citep[e.g.,][]{09guerlet, 10nixon, 18melin, 14sinclair}.  In the troposphere, the ratio of the two spin isomers of hydrogen (ortho- and para-H$_2$) can also reveal evidence for uplift \citep{82massie} - in the troposphere, abundances of para-H$_2$ that are much lower than the expectations of chemical equilibrium (sub-equilibrium conditions) imply vertical mixing from deeper, warmer levels.  Similarly, abundances that exceed equilibrium expectations imply downward mixing from higher, cooler levels (super-equilibrium conditions).   The equators of Jupiter and Saturn display sub-equilibrium conditions \citep{98conrath, 16fletcher, 17fletcher_sofia}, with weaker evidence for subsidence in the extra-tropics and polar regions.  On Uranus and Neptune, the mid-latitude display sub-equilibrium (upwelling) conditions, with subsidence at the equator and poles \citep{98conrath, 14fletcher_nep, 15orton}.  Finally, tropospheric species that should be chemically quenched at great depth (PH$_3$, AsH$_3$, GeH$_4$) are actually observed in the upper tropospheres of Jupiter and Saturn \citep[see reviews by][]{04taylor,09fouchet,18fletcher_book}.  Their spatial distributions are extremely complex, but offer tantalising hints that overturning circulation might differ between the upper troposphere \citep[where PH$_3$ is enhanced at the equator on both Jupiter and Saturn,][]{09fletcher_ph3}) and mid-troposphere \citep[where PH$_3$ and AsH$_3$ appear depleted at the equator, at least on Saturn,][]{11fletcher_vims}.  We shall return to these seemingly paradoxical gaseous distributions in the following sections.

\subsection{Lightning}

In the discussion of gaseous species above, we neglected to describe the most important volatile species for gas giant meteorology - water.  Observationally, any detection of gaseous water in the troposphere is extremely challenging \citep[e.g.,][]{04roos-serote}, and inferences of the properties of the water cloud (liquid and solid phases) are hampered by overlying cloud decks except in unique locations \citep{15bjoraker}.  Nevertheless, the morphology of plume and storm clusters on Jupiter and Saturn is reminiscent of moist convective activity, rising from the base of the water cloud into the stably-stratified upper troposphere.  And the clincher for the active role of a water cycle was the detection of lightning on both gas giants, primarily in Jupiter's cyclonic belts \citep{99little}, with significant increases in the energy of the detected lightning sferics (radio signals from the discharge) at mid-to-high latitudes \citep{18brown}.  Lightning requires powerful updrafts and the separation of charge, primarily associated with water ice, liquid, and vapour \citep[e.g.,][]{17aplin}. 

Lightning has been detected as optical flashes illuminating the clouds from below \citep{99little, 00gierasch, 07baines, 13dyudina}, radio emissions \citep{07dyudina, 11fischer}, and microwave sferics \citep{18brown}.  In addition, powerful plumes that are morphologically similar to Earth's mesoscale convective complexes (MCCs) and their associated thunderstorms have been observed in Jupiter's belts, but are either obscured or absent in the zones.  Plumes are often, but not always, associated with pre-existing cyclonic features embedded within the belts \citep{17fletcher_seb}, suggesting that cyclonicity promotes moist convective activity \citep{89dowling_dps, 16thomson}.    Saturn's lightning is very rare, but storm activity has been regularly observed at the peak of the $35^\circ$S westward jet (`storm alley'), and in the corresponding westward jet in the north.  For example, an enormous storm system erupted in Saturn's northern mid-latitudes in 2010 \citep{11fletcher_storm, 11sanchez}, with significant lightning activity correlated with the $~6$-month-long convective phase of the storm \citep[e.g.,][]{11fischer}.  The lightning distribution is subtly different to Jupiter's preference for lightning within cyclonic belts, but \citet{13dyudina} noted that lightning seemed to occur within cyclonic shear regions on small scales within the giant storm, consistent with the idea that cyclonicity promotes moist convection.  The implications for the distribution of lightning will be discussed in Section \ref{lowercell}.  
    
\subsection{Summary}

In this review so far, we have focused on the observed properties of the belts and zones of the gas giants, making some generalisations to the ice giants.  Differences have been noted between the tropics (the equator and neighbouring belts) and the Ferrel-like extra-tropics. Furthermore, we have seen that the latitudinal variations of winds, temperatures, gaseous composition, aerosols, and lightning storm activity do not always tell a consistent story.  In the next two sections, we shall describe how these might be revealing the presence of stacked meridional overturning cells, one in the stably-stratified upper troposphere (Section \ref{topcell}) and one in the mid-tropospheric weather layer (Section \ref{lowercell}), both above the deep, dry-convecting lower troposphere.

\section{Zonal Wind Decay:  An Upper-Tropospheric Cell?}
\label{topcell}

The atmosphere above the topmost condensate cloud decks (ammonia ice on Jupiter and Saturn) is readily observable in the ultraviolet, visible, and infrared.  It comes as no surprise, then, that circulation regimes in this upper tropospheric cell were the first to be explored, and informed our views of giant planet meridional overturning for the latter part of the 20th century.  Observations of both Jupiter \citep{83conrath, 98conrath, 06simon, 16fletcher_texes} and Saturn \citep{98conrath, 07fletcher_temp, 09read} have consistently revealed that the equatorward flanks of eastward jets (regions of anticyclonic vorticity) are colder than the poleward flanks of eastward jets (regions of cyclonic vorticity) throughout the upper troposphere, from the top-most clouds near 700-1000 mbar, to at least the tropopause near 100 mbar.  There is a strong correspondence between the latitudes of the zonal jets and the locations of the strongest meridional temperature gradients, such that the jet locations are co-located with the maximum vertical windshear (Fig. \ref{temp}).  This is a natural consequence of geostrophy and thermal wind balance \citep{04holton}, and the measured temperatures imply that the cloud-top winds, whether they are eastward or westward, must decay with altitude in the upper troposphere \citep[e.g.,][]{90conrath}.  Zones therefore lose their anticyclonic vorticity with height (and similar for the cyclonic vorticity of the belts) because the winds weaken with altitude.

In a stably-stratified atmosphere, the potential temperature (or, equivalently, the entropy) increases with altitude, so that rising air parcels will advect lower potential temperatures upwards, and subsiding parcels will carry higher potential temperatures (e.g., high-entropy air) downwards.  In the absence of additional heat sources, a low temperature therefore implies upwelling and adiabatic cooling, high temperatures suggests subsidence and adiabatic warming \citep{04ingersoll}.  This results in hot air sinking and cold air rising, a thermally-indirect circulation that stores potential energy and is mechanically driven. 

\textit{But what could be the source of the zonal wind decay with altitude?}    Previous works have suggested that this must be explained by a mechanical (i.e., dynamical) process: small-scale zonal eddy processes and waves providing some dissipation (a drag force) acting on the zonal flows \citep[e.g.,][]{86gierasch}, and balanced by the Coriolis force acting on the meridional circulation - i.e., a meridional circulation in balance with the drag force.  The implied mass balance from such a drag, or frictional stress, is to have rising motion equatorward of the eastward jets (e.g., upwelling and adiabatic cooling in zones); meridional flow polewards across the eastward jet (i.e., from the zone to the belt); and convergence, sinking, and adiabatic warming poleward of the eastward jets (e.g., in the belts).  


\begin{figure*}
\begin{centering}
\centerline{\includegraphics[angle=0,scale=.50]{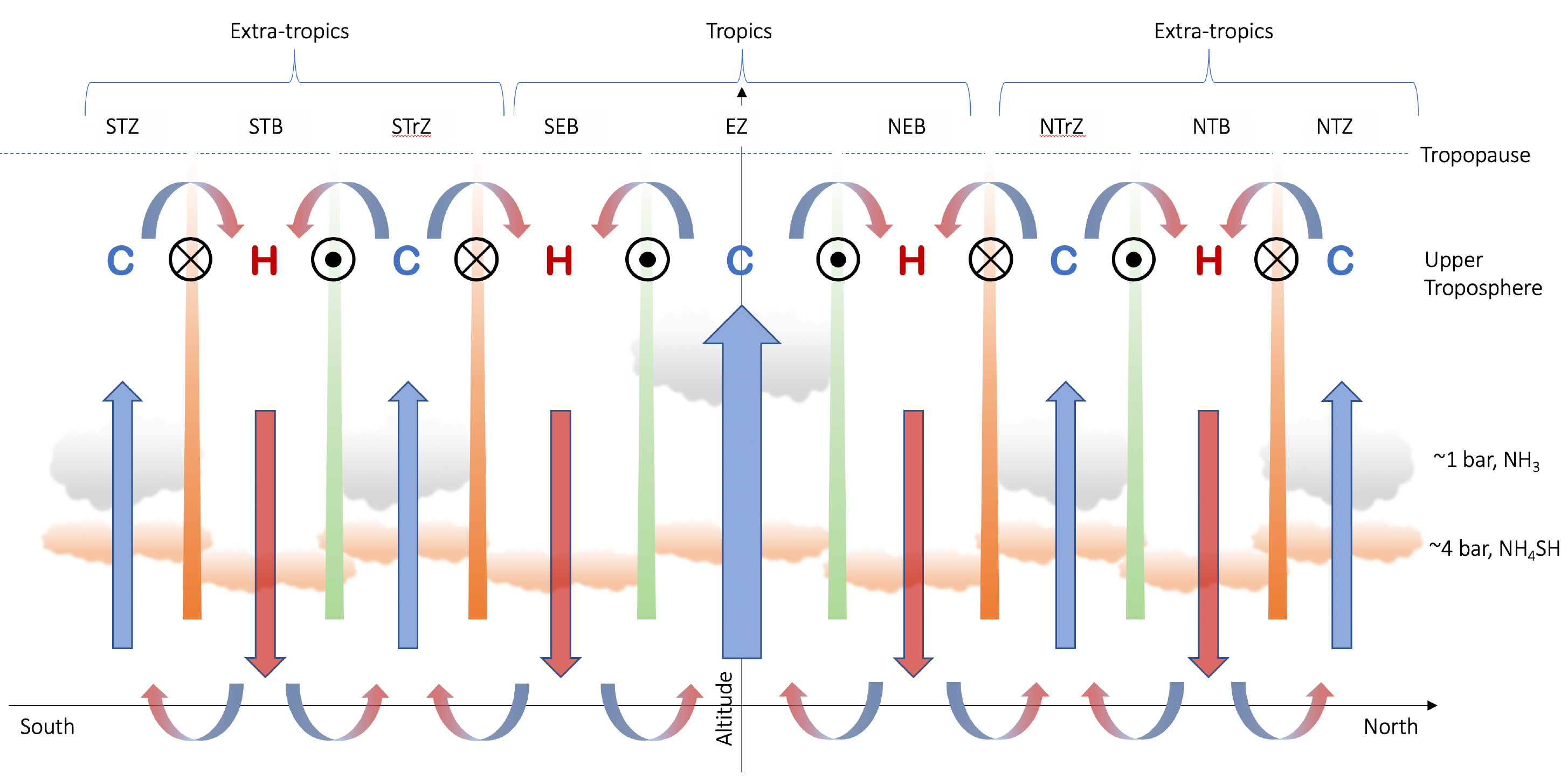}}
\caption{Schematic representation of the upper tropospheric circulation in gas giant atmospheres.  Approximate cloud altitudes and belt/zone naming conventions are shown for Jupiter.  Eastward jets are shown as green bars, with circles with black dots indicating flow `out of the page.'  Westward jets are shown as orange bars, with circles and crosses indicating flow `into the page.'  The winds decay as a function ot altitude to at least the tropopause.  The temperatures at the top of the circulation cell, as measured in remote sensing data, are indicated by `C' (cold) and `H' (hot) symbols.  Upwelling and zone-to-belt meridional circulation are shown by blue arrows; subsidence is shown by red arrows.  This can be considered as the `classical' view of belt-zone circulation, following \citet{76stone}.  }
\label{upper_cartoon}
\end{centering}
\end{figure*}

This is shown schematically in Fig. \ref{upper_cartoon} for an idealised gas giant planet, following the nomenclature for Jupiter, although it could be equally generalised to Saturn.  We depict the tropical bands (the equatorial zone and neighbouring belts, the NEB and SEB), as well as the extra-tropical bands (the northern and southern tropical zones, NTrZ and STrZ; northern and southern temperate belts, NTB and STB; and northern and southern temperate zones, NTZ and STZ).  Fig. \ref{upper_cartoon} deliberately omits processes at work below the top-most clouds, and shows what we might to consider the ``canonical'' belt-zone picture of \citet{51hess, 76stone}, with air rising and cooling in moist, cloudy zones (high-pressure centre), and sinking and warming in dry, cloud-free belts (low-pressure centres).  

The source of the jet deceleration, acting in the opposite direction to the jets themselves, remains unclear.  In the early work of \citet{86gierasch} and \citet{90conrath}, a Rayleigh friction term was used to parameterise eddy processes acting on the zonal flow.  But the underlying physical processes for these `decelerating' eddy fluxes are still debated - they could be large-scale instabilities on the jets in the upper troposphere, transferring energy from the jets into eddies \citet{89pirraglia, 93orsolini}, or some form of wave drag from gravity wave breaking \citep{04ingersoll}.  The resulting eddy stress acts to oppose the zonal winds in the upper troposphere, at least from the $\sim1$-bar clouds up to the tropopause.  However, no such processes has emerged naturally in General Circulation Models \citep[GCMs, e.g.,][]{08lian,10lian,09schneider, 18young, 20spiga}, which all tend to lack the decay of the zonal jets with altitude.  Furthermore, \citet{18showman} point out there there are no direct observations of this eddy deceleration, only an indirect inference from the decay of the zonal winds with altitude, so aside from the measured temperatures, \textit{what other observational evidence do we have from Section \ref{obs} to support the pattern of meridional circulation in Fig. \ref{upper_cartoon}?}

\subsection{Chemical Tracers of Vertical Motion}

Inferring vertical motions from chemical tracers requires a knowledge of their vertical distribution - if a species decreases in abundance with altitude (e.g., tropospheric condensables and disequilibrium species like PH$_3$), then an enhancement in a specific location implies enhanced mixing (e.g., upwelling, advection, or diffusion) from below, and vice versa for a depletion in abundance.  Conversely, if a species abundance increases with altitude (e.g., the fraction of para-H$_2$ up to its maximum at the tropopause cold-trap, or stratospheric hydrocarbon abundances increasing in altitude towards their source regions at microbar pressures), then enhanced upward mixing will cause a depletion, whereas subsidence would cause an enhancement.  However, this simple scheme ignores horizontal motions, such as large-scale meridional overturning or diffusion, which could also redistribute chemical tracers.  Caution must always be exercised when inferring vertical motions from the literature, often because the vertical distributions of these species are not well known \textit{a priori}, latitudinal mixing is poorly understood, and because spectral inversion algorithms (used to infer abundance profiles from measured spectra) are fundamentally degenerate, with the potential to confuse temperature changes, vertical distributions, and aerosol effects with real changes in a chemical tracer at a particular altitude.  Unfortunately, the historical trend is for different spectral ranges to be explored in isolation - for example,  ammonia in the mid-infrared (10 $\mu$m) has been explored and published separately from the near-infrared (2-5 $\mu$m) and sub-millimetre and radio (100 $\mu$m to 1 cm).  Believable three-dimensional distributions of these species must be able to replicate all of these spectral ranges simultaneously to be viable.

In the tropics of Jupiter and Saturn, ammonia is a particularly good tracer as its abundance should decrease with altitude due to formation of the NH$_4$SH cloud, condensation to NH$_3$ ice, and photochemical conversion to hydrazine (N$_2$H$_4$) in the upper tropospheric hazes.  In the upper tropospheric cell, infrared spectroscopy indicates upwelling in a narrow equatorial zone (latitudes less than $\sim7-8^\circ$) and subsidence in equatorial belts (latitudes of $\sim8-20^\circ$) in both hemispheres on both planets \citet{06achterberg, 11fletcher_vims, 16fletcher_texes}.  This is consistent with a Hadley-like circulation.  Such a pattern must extend into the mid-troposphere, discussed in Section \ref{lowercell}:  \citet{11fletcher_vims} and \citet{13janssen} detected the equatorial contrasts in Saturn's 2-4 bar region in Cassini 5-$\mu$m and centimetre-range remote sensing, respectively; \citet{16depater} and \citet{17li} indicated that Jupiter's equatorial region was significantly enriched in NH$_3$, down to at least 350 km of depth \citep{17li, 17bolton}.  Depending on assumptions about the equatorial aerosols and temperatures, ammonia is found to be near or exceed its saturation value at the equator, but is sub-saturated over the equatorial belts.

Additional evidence for the strong equatorial circulation comes from phosphine and para-hydrogen, two disequilibrium tracers whose vertical distribution is controlled by a balance between vertical mixing and chemical destruction.  Phosphine is enhanced at the equators of both Jupiter and Saturn \citep{04irwin, 07fletcher_ph3, 09fletcher_ph3, 16fletcher_texes}, and depleted over the neighbouring belts.  Para-hydrogen displays sub-equilibrium conditions (i.e., a decrease in the fraction of the para-H$_2$ spin isomer compared to the expectations of thermochemical equilibrium) at the equators of both worlds, indicative of uplift \citep{84conrath, 98conrath, 16fletcher, 17fletcher_sofia}.  

The discussion above pertains to the tropical regions of Jupiter and Saturn, where a Hadley-like cell may reside.  But what about the extra-tropics?  Although all sources of spectroscopic data, from near-infrared reflectivity to thermal emission in the mid-infrared, sub-millimetre and radio, show fine-scale latitudinal banding, the corresponding contrasts in chemical tracers are too small to be robustly quantified with respect to the noise.  So whilst there are tantalising hints of latitudinal maxima and minima in the chemical distributions away from the tropics, these are not on as sure a footing as the tropical contrasts.  For example, Cassini/VIMS \citep{11fletcher_vims} and RADAR \citep{13janssen, 13laraia} suggest extra-tropical variations of NH$_3$ abundance at Saturn's mid-latitudes; ground-based mid-IR spectroscopy \citep{16fletcher_texes} and centimetre-wave spectroscopy \citep{16depater, 19depater_vla} imply elevated ammonia in Jupiter's extratropical zones (NTrZ/STrZ) compared to the belts (NTB/STB); and Cassini/CIRS suggests extra-tropical variations of PH$_3$ on both Jupiter and Saturn \citep{09fletcher_ph3}.  Furthermore, Juno's microwave measurements hint at structure in the NH$_3$ distribution down to hundreds of kilometres below the clouds \citep{17li}.  However, directly assessing contrasts in chemical tracers between extra-tropical belts and zones requires high-resolution, high-signal observations, something which remains at the edge of feasibility despite the successes of Juno and Cassini.

Finally, a brief note on the use of upper tropospheric aerosols (both condensate clouds and hazes) to diagnose vertical motions.  It is important to note that uplift may not be strictly required to produce the observed increases in aerosol optical thickness in the anticyclonic zones.  Instead, the presence of thicker clouds may be a natural consequence of volatile condensation in regions of cooler temperatures, referred to by \citet{14palotai} and others as a `primary' circulation phenomenon.  So thick clouds need not imply local upwelling. However, `secondary' circulations associated with these temperature contrasts may support and enhance the contrast between the dry, aerosol-depleted belts and the moist, aerosol-covered zones, and indeed some mechanism is required to re-supply condensed volatiles that are removed from the upper troposphere via precipitation \citep[e.g.,][]{17ingersoll, 19li}.  In the tropics, the correspondence between cool zones and cloudiness is clear.  However, in the extra-tropics of Jupiter and Saturn, cyclonic belts that are warm in the upper troposphere are not necessarily cloud-free \citep{19antunano} - the 5-$\mu$m bright bands in Jupiter's extra-tropics are not perfectly co-located in the cyclonic regions (with eastward jets on their equatorward sides, and westward jets on their poleward sides).  All this suggests that \textit{the presence or absence of aerosols does not imply vertical motion}, even though vertical motion would enhance aerosol contrasts.

Fig. \ref{upper_cartoon} summarises our present knowledge of gas giant bands in the upper troposphere, via upwelling in cool zones, subsidence in warm belts, deceleration of zonal winds with altitude due to some poorly-constrained eddy friction, and elevated chemical tracers (and to a lesser extent aerosols) in the anticyclonic zones.  But, as we shall see in Section \ref{lowercell}, this `classical view' is by no means the end of the story.


\section{Eddy-Driven Jets:  A Middle-Tropospheric Cell?}
\label{lowercell}

The `classical view' of giant planet meridional circulation was challenged in the past two decades by new findings from the Galileo and Cassini missions - namely the localisation of lightning activity and the observations of how the zonal jets are maintained.  Here we describe the observational evidence that an eddy force at cloud-level is serving to accelerate the zonal jets, in the opposite sense to the zonal jet deceleration happening in the upper cell.  Small-scale eddies and turbulence are generated by a variety of processes in the weather layer, including thunderstorms in moist-convective complexes, and natural instabilities arising from gradients in the temperatures and zonal wind flow (barotropic instabilities from horizontal shears, baroclinic instabilities from vertical shears).  The energy for these eddies could come from the planet's internal heat, or from the potential energy stored within the thermal gradients themselves, manifesting as instabilities.

Building on explorations of eddy fluxes by Voyager \citep{81ingersoll}, cloud-tracking from the Cassini spacecraft provided evidence that eddy momentum was converging and supplying momentum to the eastward zonal jets on both Jupiter and Saturn \citep{06salyk, 07delgenio, 12delgenio}, implying an upscale transfer of energy from the smallest (eddies) to the largest (zonal) scales.  This conclusion required accurate and precise measurements of the eddy velocities via automated image correlation techniques in order to measure the stress exerted on the jets \citep{09delgenio, 18showman}.  As the eddy motions are generally weak, these calculations can be noisy, but show significant correlation between the eddy momentum flux and the variation of zonal velocity with latitude at cloud level.  Jet forcing by the convergence of eddy momentum fluxes is analogous to Earth's Ferrel cells \citep{07hartmann}, and is a characteristic of baroclinic instability in Earth's mid-latitudes \citep{47charney, 49eady, 82pedlosky}.  But this terrestrial analogy may not be entirely relevant, given that this has not been directly observed, even though baroclinic instabilities are theoretically possible on gas giants \citep{07kaspi, 08lian}, and are often produced in numerical simulations \citep{09delgenio}.  Indeed, an analysis of Jupiter's kinetic energy spectrum by \citet{17young} suggested that the jet forcing mechanism could have a dominant energy source at the appropriate length scales (i.e., the deformation radius) for baroclinic instabilities.

\begin{figure*}
\begin{centering}
\centerline{\includegraphics[angle=0,scale=.50]{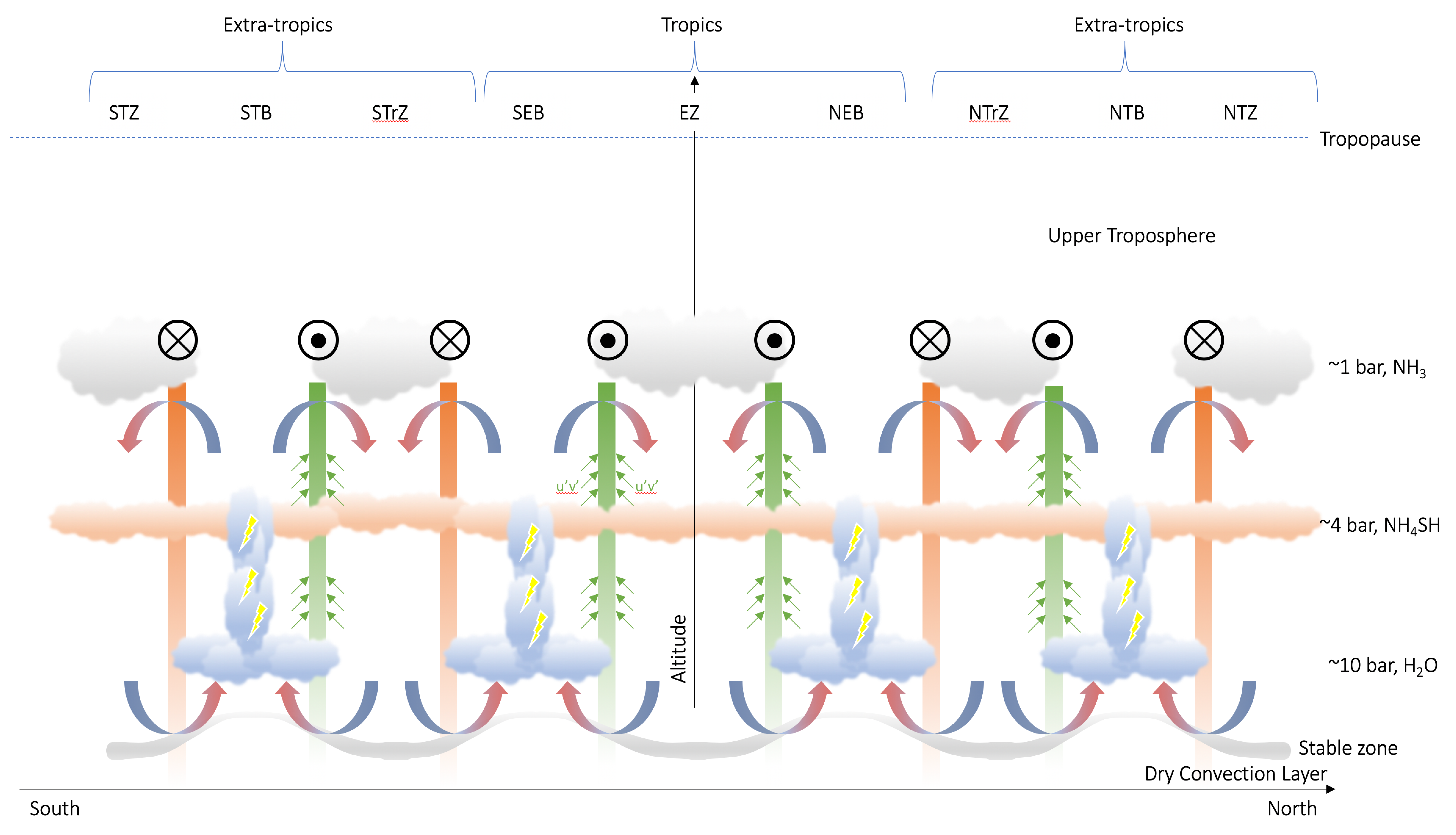}}
\caption{Schematic representation of the mid-tropospheric circulation in gas giant atmospheres.  Approximate cloud altitudes and belt/zone naming conventions are shown for Jupiter.  Eastward jets are shown as green bars, with circles with black dots indicating flow `out of the page.'  Westward jets are shown as orange bars, with circles and crosses indicating flow `into the page.'  Small green arrows indicate eddy momentum flux into the eastward jets; plume activity with putative lightning are indicated in the centres of the cyclonic belts; and associated belt-to-zone meridional transport is indicated that the top of this cell.  The stable inversion layer beneath the water cloud, separating the moist weather layer from the dry convection lower troposphere, is indicated by the grey shaded line.  This is thinner beneath cyclonic belts, and thicker beneath anticyclonic zones, following \citet{16thomson}.  This schematic view echoes that envisaged by \citet{09delgenio} (their Figure 6.18).}
\label{lower_cartoon}
\end{centering}
\end{figure*}

This forcing by flux convergence must be balanced by another force in the zonal direction.  For small Rossby numbers, the prevailing zonally-averaged balance in the zonal direction is between the eddy-forcing and the Coriolis acceleration on the associated meridional circulation, and this is found in numerical simulations \citep[e.g.,][]{08lian, 20spiga}.  The meridional force balance is a geostrophic balance between the Coriolis force and the pressure gradient, leading to the standard thermal wind equation.  As pointed out by \citet{18showman}: ``it is generally incorrect to think of thermal-wind balance and eddy driving of zonal jets as being either-or," they are part of the same process.  At the top of this mid-tropospheric cell, the meridional motion must balance Coriolis accelerations against the eddy angular momentum flux convergence, whereas at the bottom the lower branch would balance Coriolis accelerations against any drag forces at work at depth.  The resulting circulation requires rising motions poleward of the eastward jets (i.e., in the belts); equatorward meridional flow across the eastward jets (e.g., from the belt to the zone), and sinking motion equatorward of the eastward jets (i.e., in the zones) \citep{07showman, 08lian, 09delgenio, 10liu}. This belt-to-zone circulation is summarised in Fig. \ref{lower_cartoon}, and is the complete opposite of that described in the upper troposphere (Section \ref{topcell} and Fig. \ref{upper_cartoon}).  


The meridional motions in this mid-tropospheric cell might help to explain the preferential occurrence of lightning within the cyclonic belts, as first postulated by \citet{00ingersoll} and described in Section \ref{obs}.  If the lower cell extended to the water clouds, it would imply moist air converging and rising in the belts, as required for the existence of lightning.  But does this imply large-scale net rising in the belts, or more localised ascent in thunderstorms, punching into a region of overall subsidence?  Net belt-wide ascent seems unlikely, as it would elevate aerosols and condensables throughout the belt, opposite to what is observed.  In particular, the absence of a global water cloud layer in the belts would appear inconsistent with the idea of considerable quantities of water being added to the belts.  Furthermore, the addition of substantial amounts of moisture to the belts would generate a moist adiabat and reduce CAPE \citep[convective available potential energy,][]{94emanuel}, ultimately inhibiting the thunderstorms.  Indeed, the presence of thunderstorms in the belts requires that belts have high values of CAPE \citep{05showman}.  However, is this small-scale rising in convective events in cyclonic belts sufficient to explain the belt-to-zone meridional circulation needed to balance the eddy-forcing of the jets?  Or is belt-wide convergence at the depths of the water clouds truly required to create large reservoirs of CAPE within the belts? This additional conundrum has not yet been studied in any depth.

Maybe the existence of lightning does not actually require net upwelling in the belts, but is a result of the cyclonic vorticity and the implied lower pressures, which suggests an upward bulge of the boundary between the dry-convecting lower troposphere and the moist mid-troposphere.  Moist convection is peculiar on giant planets, as the condensates (e.g., water molecular mass of 18) are always heavier than `dry air' (e.g., mean molecular weight of 2.2 g/mol on Jupiter and Saturn), leading to potential suppression of moist convection \citep{95guillot, 17leconte, 17friedson}. As the loading of heavy volatiles into H$_2$-He dominated air has a stabilising effect, a stable interface is created between the deep, dry-convective lower troposphere, and the moist mid-tropospheric weather layer \citep{14sugiyama, 15li}.  This stable inversion layer can actually inhibit moist convection until CAPE has accumulated to some critical level.  \citet{89dowling_dps} suggested that the upward bulge of the boundary implies that the stable layer is thin beneath both belts and smaller-scale cyclonic features, allowing air to rise from the dry-convecting depths to the lifting condensation level \citep{16thomson, 17fletcher_seb}, triggering convective outbursts that relinquish CAPE.  Indeed, on small scales, such as within the cyclonic `barges' in Jupiter's belts \citep{89dowling_dps, 17fletcher_seb}, or the cyclonic regions associated with Saturn's large-scale storms \citep{18sanchez_storm}, cyclonicity appears to promote moist convection and, presumably, lightning.  This ``charge-recharge'' cycle for water may partially explain the observed multi-year cycles in Jupiter's belts \citep{17fletcher_cycles} and Saturn's storms \citep{18sanchez_storm}.  Conversely, the downward bulge beneath anticyclonic zones would cause the stable layer to be so thick and deep that thunderstorms are effectively inhibited \citep{05showman}.  So in this scenario, the belts could still be regions of net subsidence, but their cyclonicity and large values of CAPE promotes the moist convection.

\textit{Besides the observations of eddy-flux convergence and lightning, are there other suggestions that might support this reversed meridional overturning in the weather layer?}  Evidence for a different circulation pattern can be tentatively found in Saturn's distributions of arsine and phosphine derived from Cassini/VIMS data in the 1-3 bar range \citep{11fletcher_vims}.  These distributions are the opposite to the NH$_3$ distribution inferred at similar altitudes, and to the PH$_3$ distribution inferred in the upper troposphere from Cassini/CIRS \citep{09fletcher_ph3}, raising a conundrum.  As the chemical transformations responsible for depleting arsine and phosphine occur at deeper levels (kilobar pressures), the equatorial depletions (and enhancements in the neighbouring belts) could be indicative of a reverse circulation in the 1-3 bar range. \citet{17giles} used high-resolution $5-\mu$m spectroscopy to explore whether the same was true on Jupiter, but found that the data were consistent with a latitudinally-uniform distribution of germane, and with phosphine and arsine increasing from equator to poles.  Their study highlighted the potential degeneracies associated with the fitting of the 5-$\mu$m spectra:  although the spectral inversion algorithm did display contrasts between the equatorial zone and neighbouring belts that might support a reversed circulation in the 1-3 bar region of Jupiter, the change in the goodness-of-fit to the spectra (compared to simply holding the species constant with latitude) was too small to be considered robust.  Until the degeneracy between aerosols and composition can be robustly broken, the inferred circulations from 5-$\mu$m spectra must be treated with extreme caution.  Furthermore, despite evidence for equatorial minima in arsine and phosphine, it is very hard to reconcile this with the equatorial maximum in NH$_3$ on both Jupiter \citep{17li, 16depater} and Saturn \citep{11fletcher_vims, 13janssen}, which suggests upwelling over great depths at the equator.

The eddy momentum convergence cannot be occurring over a vast vertical column, deep into the atmosphere.  \citet{06salyk} found the rate of conversion of eddy to zonal mean kinetic energy was about 4-8\% of Jupiter's thermal energy flux, assuming that the weather layer was $\sim2.5$ bar deep.  Furthermore, \citet{12delgenio} measured about half this energy conversion rate for Saturn based on similar assumptions.  The eddy pumping must therefore be vertically localised, a shallow process - not to be confused with a statement on the ultimate depth of the zonal winds, which can be much deeper into the dry convecting region \citep{18kaspi, 18guillot, 19galanti, 19iess}.  \citet{06showman} and \citet{08lian} explicitly demonstrated in analytical and numerical models that jet forcing confined to shallow levels of a few bars or less can easily lead to zonal jets that penetrate far deeper than the level of the jet forcing - this is a process known as `downward control' and also happens on Earth \citep{91haynes}.  We might reasonably expect the jets to be fastest at the altitudes where this eddy pumping is most efficient, and to then decrease above this altitude (into the upper troposphere) and below this altitude (into the dry convecting lower troposphere) due to an unspecified and unidentified damping process.  Indeed, \citet{12delgenio} inferred that the eddy flux convergence weakened with height in Saturn's upper troposphere, suggesting that the processes responsible for the eddies are driven from the mid-troposphere (e.g., via instabilities) rather than from above (e.g., radiative processes in the hazes).  Viewed from another perspective, might this imply that we could detect a region, just below the topmost clouds, where winds are strengthening with altitude to the point where eddy pumping is most efficient?  

Recent evidence for such a vertical strengthening of winds has been identified on Saturn and Neptune.  \citet{18studwell} suggested that zonal winds in Saturn's extratropical regions actually strengthened from 2 bars (sensed by Cassini/VIMS at 5 $\mu$m) to 0.5 bars (sensed by Cassini/ISS in the visible), although we caution the reader that the differences identified were small, particularly compared to Saturn's tropics where the winds weaken with increasing altitude.  For wind speeds to decay with increasing depth into the troposphere, thermal wind balance would imply temperature gradients \textit{in the opposite sense} to those observed in the upper troposphere - warm zones and cooler belts.  

On Neptune, Voyager/IRIS and Earth-based mid-infrared observations reveal cool mid-latitudes and a warm equator and pole, suggesting that Neptune's strong retrograde jet should weaken with altitude in the 80-800 mbar range \citep[e.g., an upper tropospheric cell above the clouds,][]{98conrath, 14fletcher_nep}.  Thus the winds were expected to strengthen with depth into Neptune's lower troposphere.  However, \citet{18tollefson} showed that low-latitude cloud tracers observed in the H-band (sensing the deep clouds at $p>1$ bar) were retrograding more slowly than those observed in the K band (sensing lower stratospheric clouds near 10 mbar), implying a weakening of the westward flow with depth.  This in turn would imply mid-troposphere meridional temperature gradients \textit{in the opposite sense} to those found from Voyager in the upper troposphere.  

We stress that the reversals of the temperature anomalies, from cool zones to warm zones with increasing depth, have never been directly observed.  They are only inferred from small differences in observed windspeeds as a function of depth. As an interesting comparison, numerical simulations of prominent anticyclones like Jupiter's Great Red Spot \citep{14palotai} expect to see warm cores at altitudes below the 500-mbar pressure level (the `mid-plane' of their vortex simulations).  Observations suggest that this warm core must actually be deeper ($p>1$ bar) so that it is not detectable in thermal-infrared observations, and indeed recent observations by Juno's microwave instrument have suggested a warm region beneath the Great Red Spot at high pressures \citep{17li_grs}.  Maybe the idea of a `mid-plane' also applies to the belt/zone structure, and the mid-plane marks the transition into the mid-tropospheric cell shown schematically in Fig. \ref{lower_cartoon}.

\section{The Multi-Tier Circulation Cell Concept}
\label{two-cell}

\textit{Can we really divide giant planet atmospheres into circulation regimes with opposing meridional flows?  Or are we simply misinterpreting the observational data, imagining a paradox that does not truly exist in reality?}  We saw in Section \ref{topcell} that the `classical view' of upwelling in cool, anticyclonic zones, with zone-to-belt mass transfer via a meridional circulation, was a perfectly adequate representation of the observed temperatures, zonal-wind decay with altitude, and the majority of gaseous and aerosol tracers.  In Section \ref{lowercell}, we showed that upwelling in the cyclonic belts, at the depth of the water cloud, could explain the distribution of lightning, and the belt-to-zone meridional transport was sufficient to balance the eddy momentum flux convergence in the eastward jets.  If eddy processes are truly the culprit, then we must explain how there can be a transition from eddy acceleration (forcing the jets at the clouds and below) to eddy deceleration (causing jets to decay with altitude), and where such a transition, or `mid-plane,' could occur in the atmosphere.  A researcher's interpretation of banded structure in remote sensing observations will depend on whether their data senses above or below this transitional level.  

\begin{figure*}
\begin{centering}
\centerline{\includegraphics[angle=0,scale=.50]{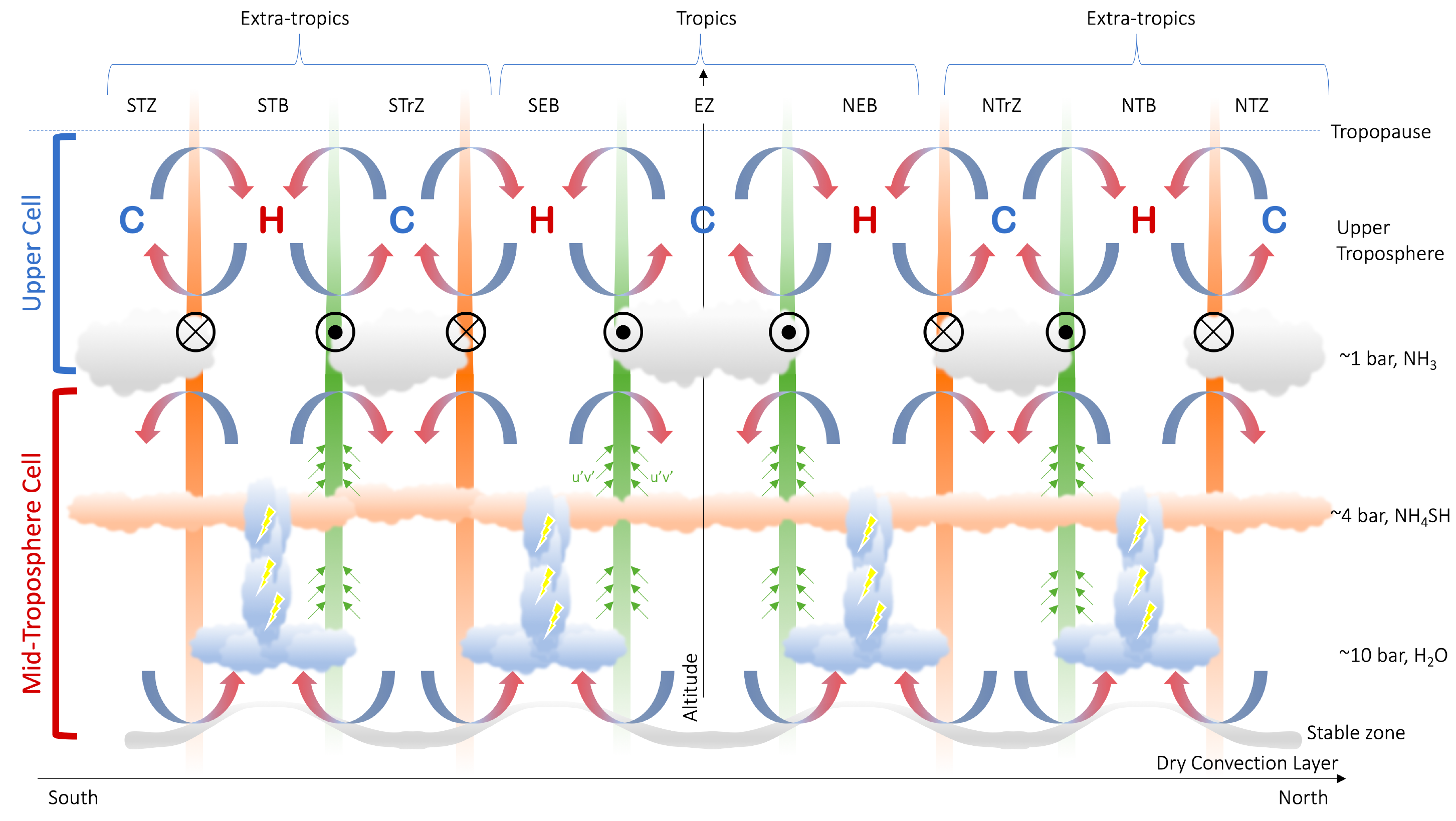}}
\caption{Schematic representation of stacked cells within a gas giant atmosphere, following the same naming conventions as those in Fig. \ref{upper_cartoon} and \ref{lower_cartoon}.  Strong equatorial upwelling, penetrating throughout both cells, is indicated by the thick blue arrow.  At the transition point (or `mid-plane'), placed somewhat arbitrarily near the top-most clouds, we would have divergent flow over belts and convergent flow over zones.  This schematic expands upon that first shown in \citet{05showman} and described by \citet{00ingersoll}.  Note that this picture lacks the deep equatorial upwelling that might - or might not - be required to explain the enhanced equatorial NH$_3$ (see main text).   Can reality truly be this complex?}
\label{stacked_cartoon}
\end{centering}
\end{figure*}

As described by \citet{00ingersoll}, ``when both views are considered together, they imply that, in the belts, the two vertical currents converge within the clouds.''  We therefore combine our schematics for the upper and mid-tropospheric cells in Figs. \ref{upper_cartoon}-\ref{lower_cartoon} to generate a stacked cell schematic in Fig. \ref{stacked_cartoon}.  This schematic echoes previous discussions of multi-tier stacked circulation cells by \citet{00ingersoll}, \citet{05showman}, \citet{11fletcher_vims} and \citet{18showman}.  We place the transition point at the cloud tops, immediately above the altitude where the cloud tracers imply eddy momentum flux convergence, although we stress that this is little more than an educated guess, and that this pressure could vary substantially depending on location.  If we place the transition level above the clouds, then we would be able to explain the eddy-momentum flux observations but not the condensation of the cloud bands themselves.  Conversely, if we place the transition level beneath the clouds, then we can explain the temperatures (Fig. \ref{temp}) and upper-tropospheric chemical tracers, but not the eddy-momentum flux observations.  Unfortunately this conundrum cannot be resolved in this review. 


Figure \ref{stacked_cartoon} attempts to encapsulate all the various competing evidence for meridional circulations on Jupiter and Saturn.   Zonal winds strengthen slightly with height through the mid-troposphere, then weaken above the cloud tops, broadly consistent with Galileo probe findings on Jupiter \citep{98atkinson} and Cassini measurements of Saturn's extra-tropical winds between 0.5 and 2 bar \citep{18studwell}.  It explains the pattern of warm, cloud-free, and gas-depleted belts and cool, cloudy, and gas-enriched zones measured in the upper troposphere.  And it partially explains the Cassini observations of PH$_3$, AsH$_3$ (but not NH$_3$) in Saturn's mid-troposphere \citep{11fletcher_vims}.  But both Juno \citep[e.g.,][]{17li, 17ingersoll} and Cassini \citep[e.g.,][]{11fletcher_vims, 13janssen} have revealed that the tropics are somewhat unique - NH$_3$ gas is highly enriched at the equator in the upper troposphere and mid-troposphere, potentially down to 40-60 bar.  

In the stacked cell picture, the gas-enriched air rising in the zones in the upper cell would have been primarily provided by lateral motions from the belts at the transition point, rather than being supplied directly from below \citep{05showman, 00ingersoll}.  This was useful to explain why Jupiter's ammonia was apparently depleted in the 1-5 bar range \citep{01depater}, and indeed the Juno microwave observations do suggest some kind of boundary near 10 bars where the ammonia gas depletion stopped (C. Li, pers. comms.), potentially suggesting that Jupiter's mid-tropospheric cell (away from the equator) is NH$_3$-depleted and extends down to the base of the water cloud.  But this clearly does not explain the equatorial NH$_3$ enrichment revealed by Juno and Cassini, and a reconciliation of the microwave results (suggesting equatorial upwelling) with the eddy-forced jet stream circulation (suggesting equatorial downwelling) is needed.  Maybe the contrast lies in the differences between Hadley- and Ferrel-like circulation patterns, with the former dominating the equatorial temperature, wind, and chemical tracer fields for Jupiter and Saturn \citep[e.g.,][]{05yamazaki}.  Or maybe the equatorial NH$_3$ enhancement is unrelated to upwelling, and instead related to some unforeseen processes within the clouds.


Inferring circulation patterns from remote-sensing measurements is hampered by the fact that different spectral ranges - near-infrared reflectivity, and thermal emission in the mid-infrared, sub-millimetre, and radio - are all analysed independently.  It seems somewhat convenient that Cassini/CIRS observations might be sensing Saturn's upper-tropospheric cell, whereas Cassini/VIMS observations at 5 $\mu$m are providing access to the mid-tropospheric cell.  Simultaneous inversion of spectra from multiple instruments, with particular focus on breaking the composition-aerosol degeneracy, is now required to truly understand how these gaseous species vary as a function of altitude.

Furthermore, we might reasonably ask \textit{whether we really need upwelling and downwelling to explain the observed belt/zone contrasts?}  Here we urge caution, and that each potential tracer (temperatures, gases, aerosols) be considered on their own merits.  Recent models for discrete anticyclones \citep{10depater, 13marcus, 14palotai} focus on their expected two-tiered structures, with high pressure due to convergence at a mid-plane, rising secondary circulation in the centre of an upper tier, and falling secondary circulation in the centre of the lower tier\footnote{The secondary circulation in the GRS is even more complex, as the observation of a warm, weakly cyclonic core suggests sinking motion in the very centre of the anticyclone \citep{10fletcher_grs}}.  As such, this is analogous to our discussion of convergence in the mid-plane for a zone. For example, \citet{14palotai} show that cold-core temperatures above large and thin anticyclones are `primary' circulation features:  ``the anticyclone's top centre is cool so that it can fit under the tropopause, which resists bulging up to make room.''  The cool temperatures lead to condensation of the vapours \textit{in place}, rather than requiring upwelling via a slow `secondary' circulation (i.e., one that has been driven by the thermal anomalies associated with the primary circulation).  \citet{14palotai} argue that the weak secondary circulations drive the distributions of chemical tracers, such as the distribution of phosphine, or the coloration of the large anticyclones.  Upwelling/downwelling are then not \textit{required} to explain the temperature and cloud contrasts, but secondary upwelling/downwelling might be induced by (and ultimately reinforce) those same contrasts.  In summary, a two-tiered structure is certainly viable, but the meridional circulation is a secondary and very slow consequence of the primary temperature/wind patterns.  Indeed, Earth's mid-latitude clouds are largely unrelated to upwelling and subsidence in the Ferrel cells, but are instead a consequence of the baroclinic instabilities within the cells \citep[e.g.,][]{06vallis}. 

If the eddy forcing in the mid-tropospheric cell must be truly balanced by a meridional circulation as shown in Fig. \ref{lower_cartoon}, then the idea of a tier of stacked circulation cells is still required.  \textit{So how might we explain the transition from eddy-driving of the jets in the mid-troposphere to eddy-dissipation of the jets in the upper troposphere?}  There exists a natural transition from convectively-unstable to statically-stable around the radiative-convective boundary, which exists near 300-500 mbar on Jupiter and Saturn, between the condensate clouds and the tropopause.   If the eddies emerge from moist convection or convective instabilities in the unstable adiabatic region, then this might explain why eddy momentum flux convergence is restricted to the mid-tropospheric cell, rather than in the stably-stratified upper cell.  Furthermore, wave propagation is extremely sensitive to this background stability and the curvature of the windfield, which would lead to waves and eddies behaving differently between the upper and middle tropospheres.  As waves transfer momentum to the mean zonal flow \citep[e.g., a good example being the interaction of waves with the sinking pattern of jets and temperature anomalies associated with the Earth's Quasi-Biennial Oscillation,][]{72holton, 01baldwin}, it is not inconceivable they they could be one of the underlying mechanisms responsible for the frictional drag on the zonal jets in the upper troposphere.  Indeed, the Saturn GCM developed by \citet{20spiga} shows decay of the equatorial jet with altitude because of the prevalence of westward-propagating Rossby waves near the equator, supporting the idea of wave breaking as the source of dissipation, although the details are yet to be determined (A. Spiga, \textit{pers. comms.}).  Viewed in this light, the transition between the different tiers can be considered as a consequence of the increasing importance of sunlight, generating statically-stable conditions above the clouds.

Multi-tiered structures on the scale of the belts and zones have emerged in numerical simulations in the past.  For example, the analytical meridional overturning model of \citet{09zuchowski} resulted in two counter-rotating meridional circulation cells - an upper cell associated with radiative forcing (which they label as `stratospheric') consistent with the belt/zone temperature field, and a deeper eddy-driven cell consistent with the measurements lightning and eddy-momentum flux convergence.  They showed that shifts in the direction of vertical motion in this tiered structure could be ``enforced by confinement of active forcing to a finite layer and the application of drag on the zonal jets below and above this region.''  

Using an earlier version of the same model, \citet{05yamazaki} explored the equatorial winds on Jupiter and Saturn via a combination of momentum transfer from an equatorial Kelvin wave with meridional circulation analogous to a Hadley cell.  Although the setup was somewhat contrived, their Hadley-only scenario produced eastward jets at the poleward edges of a thermally-direct Hadley cell, and identified ``a weak counter-circulation underneath the main, thermally-direct circulation'', which they suggested was a frictional response to the Hadley circulation of their upper cell.  Removing the Hadley circulation and considering only the waves, they found that the zonal jet velocities matched the phase speed of the wave at altitudes where eddy forcing was maximised and the wave dissipates, depositing energy into the mean zonal flow.  Above this altitude, the jet then decayed into the upper troposphere. In this wave-only scenario, they also showed two thermally-indirect downwelling cells at the equator.  Finally, combining both the Hadley circulation and the wave-driven flow, they reproduced the approximate Jovian and Saturnian wind structure (with eastward peaks off the equator), and identified stacked cells - thermally-direct Hadley cells in the upper troposphere, and thermally-indirect cells driven by the dissipating Kelvin wave beneath.  They concluded that the structure was determined by the fine balance between wave generation at depth and dissipative processes in the upper troposphere.

More recent general circulation models (GCMs) support the idea that eddy momentum flux and the balanced meridional circulation changes with altitude.  For example, Fig. 8 of \citet{20spiga} show a cross-section of Saturn's eddy momentum transport as an output of their model, indicating how eddy momentum convergence switches to divergence at around 200-300 mbar.  Furthermore, close inspection of their Fig. 7 suggests that mid-latitude prograde jets occur on the equatorward sides of cool zones, with retrograde jets on the poleward side, and a general increase in the zonal winds with altitude (away from equator).  These temperature contrasts and winds are all consistent with the Saturn GCM reproducing our hypothetical mid-tropospheric cell, where winds increase with altitude.  The pressure of the transition to the upper cell, which is not well reproduced in the GCM, is presumably set by the background stratification predicted by the radiative transfer scheme in the model.  

A new GCM for Jupiter by \citet{18young}, building on previous work by \citet{09zuchowski, 09zuchowski_aer}, also produces eddy-driven mid-latitude jets and the equatorial superrotation (in the case of applying an internal heat flux).  They find that the magnitude of the meridional circulation weakens significantly above 300 mbar (maybe implying a transition), and that the meridional motion is strongest at Jupiter's equator compared with the neighbouring belts, consistent with the elevated chemical tracers detected there.  Fig. 12g of \citet{18young} provides further insight into these Ferrel-like cells - the meridional circulation associated with eastward jets have divergence on their poleward side, and convergence on their equatorward side, consistent with upwelling in belts and belt-to-zone mass transfer as in our hypothetical mid-tropospheric cell (R. Young, pers. comms.).  However, they do not see any evidence for upper-level meridional circulation in the opposite sense.  

If the stacked, multi-tier pattern of meridional circulation cells exists, then the production of waves at depth, and their dissipation in the stably-stratified upper atmosphere (e.g., wave breaking providing a frictional stress), might be the key to the transition from the domain of eddy pumping to the domain of zonal-wind decay.  Intriguingly, the shifting balance between the differing meridional overturning strengths, or changes to the efficiency of wave propagation, could help to explain the observed temporal variability in the weather layer.  When a source of momentum to the zonal winds is disturbed, we might expect to see a shift in the balance of the meridional motions responsible for maintaining the banded pattern \citep[e.g.,][]{09zuchowski}.  For example, the apparent shutdown in the small-scale convective activity near the Great Red Spot heralded a transition for Jupiter's South Equatorial Belt (SEB) to a quiescent, faded phase \citep{11fletcher_fade}.  This faded and whitened state, due to the formation of high-altitude ``cirrus'' over the SEB, persisted for many months until spectacular convective storm-clusters erupted from the locations of cyclonic features within the SEB \citep{17fletcher_seb}, reinvigorating the brown appearance over the course of several months.  We might now begin to understand this change - and similar variations in the other jovian belts \citep{17fletcher_cycles} - as an altered balance between the meridional circulations of the two cells within the belt when moist convection shuts down, no longer producing the eddies and waves that maintain the circulation.

Finally, \textit{what happens in the lowest tier of the troposphere, in the dry convective region depicted in Fig. \ref{stacked_cartoon}?}  We described in Section \ref{obs} how recent results have shown that the zonal jets on Jupiter and Saturn penetrate to depths of 3000 km \citep{18kaspi} and 9000 km \citep{19galanti, 19iess}, respectively.  But this can still be consistent with jets formed by relatively shallow processes in the weather layer \citep{06showman, 08lian}.  Although the eddy pumping existed in only the top few bars around the cloud-forming layers, and zero at deeper pressures, these authors showed that meridional flow could extend deeper, giving rise to zonal jets at all depths.  Deep jets and deep belt/zone contrasts need not, therefore, have a deep origin \citep[although we make no distinction between the applicability of shallow-atmosphere or deep-atmosphere numerical simulations, e.g.,][]{05vasavada}.  Note that this also satisfies the requirement that energy conversion from eddies to the kinetic energy of the zonal flow only be happening over a relatively shallow layer (Section \ref{lowercell}), so that it does not exceed the observed internal luminosity of these worlds.

\section{Extension to the Ice Giants}
\label{large-scale}

The large quantity of Jovian and Saturnian observations, and numerical simulations, means that this review has been biased heavily towards the gas giants.  In this final section, we explore the potential implications for multi-tier stacked cells for the ice giants, Uranus and Neptune.  In Section \ref{obs} we introduced the idea that the observed banding pattern on the ice giants is different to the gas giants.  Namely, the ice giants exhibit:
\begin{itemize}
\item Broad retrograde jets at their equators and a single prograde jet at high latitudes in each hemisphere \citep{93sromovsky, 15sromovsky, 18sanchez_jets};
\item Cool mid-latitudes contrasting with a warm equator and pole in the 80-800 mbar range (Fig. \ref{temp}), implying decay of the zonal winds with altitude in the upper troposphere \citep{87flasar, 98conrath, 14fletcher_nep, 15orton};
\item Hints that the equatorial retrograde winds decline in strength with increasing depth, counter to the expectations from the thermal wind equation in the upper troposphere \citep{18tollefson};
\item Sub-equilibrium para-H$_2$ abundances at mid-latitudes, coinciding with the location of notable storm activity, suggestive of mid-latitude upwelling \citep{98conrath, 14fletcher_nep, 15orton}; 
\item Banded albedo patterns on a much finer latitudinal scale than the temperature and wind field \citep[e.g.,][]{89smith, 12fry, 01sromovsky};
\item An equator-to-pole decrease in the abundance of methane in the troposphere of Uranus \citep{09karkoschka, 14sromovsky} and Neptune \citet{11karkoschka_ch4} suggestive of large-scale motion from low- to high-latitudes - note that \citet{14sromovsky} suggest that their 0.5-1.0 $\mu$m data sense the $p>1$ bar region (their Fig. 2), below the levels sensed by Voyager's infrared instruments;
\item Microwave-bright polar regions suggestive of depletion of NH$_3$ and/or H$_2$S at high latitudes \citep{91depater, 03hofstadter, 14depater}.
\end{itemize}

\begin{figure*}
\begin{centering}
\centerline{\includegraphics[angle=0,scale=.50]{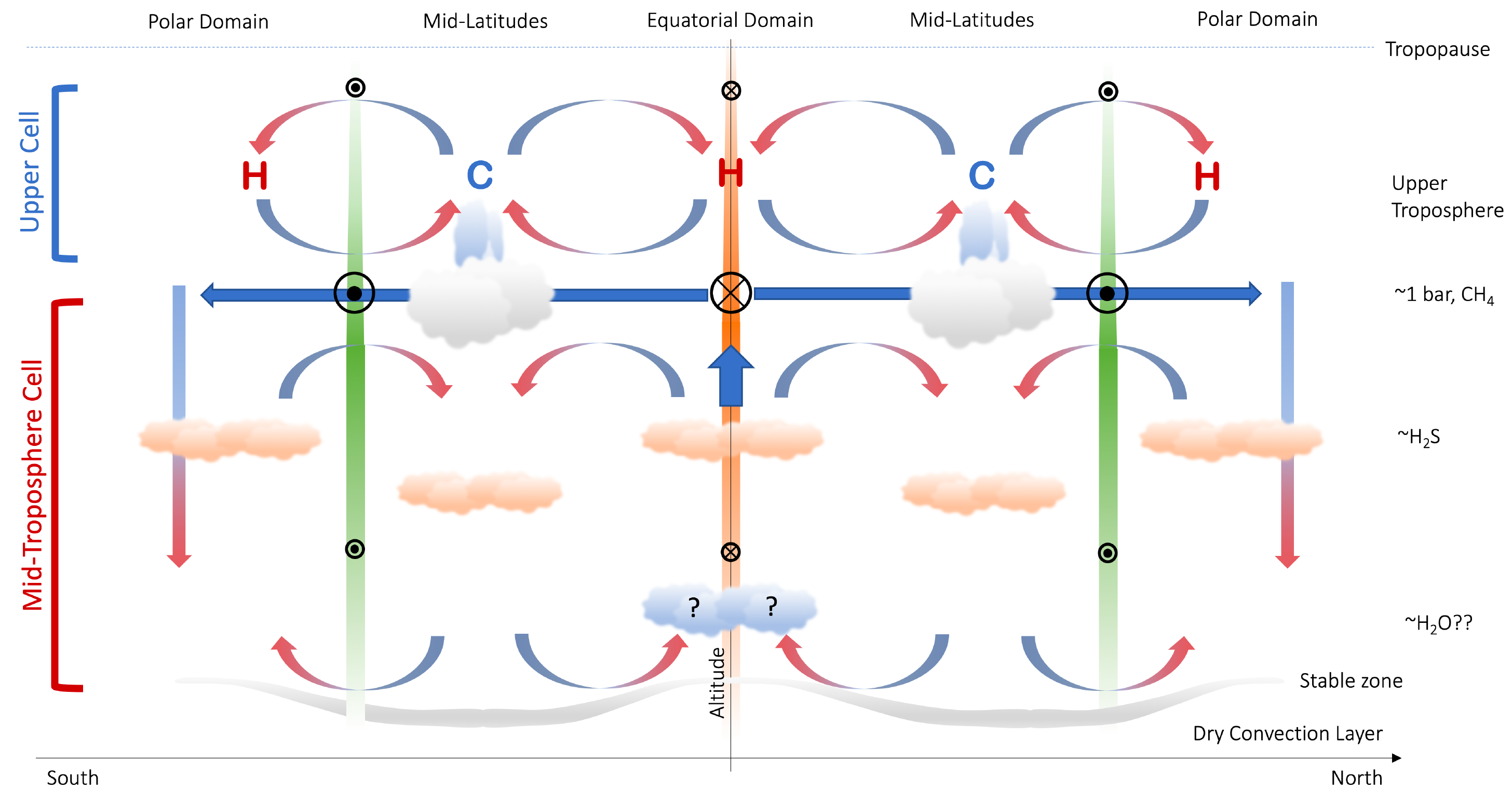}}
\caption{Schematic representation of stacked cells within an ice giant atmosphere.  The equatorial retrograde jet is shown in orange and decays both above and below the clouds \citep{98conrath, 18tollefson}.  No eddy momentum fluxes are included here.  The gradient in methane from equator to pole is shown as the thick blue arrow for $p>1$ bar, whereas the mid-latitude upwelling and equatorial/polar subsidence is shown for $p<1$ bar.  Strong polar subsidence then explains the microwave brightness and NH$_3$/H$_2$S depletion.  The variable thickness of a stable layer associated with water clouds is pure conjecture.}
\label{cartoon_icegiant}
\end{centering}
\end{figure*}

An attempt to reconcile these findings is shown in Fig. \ref{cartoon_icegiant}, and adapts similar schematics depicted in \citet{14sromovsky}, \citet{14depater} and \citet{18tollefson}.  The methane and microwave-brightness contrasts could be explained by single large-scale cells, with equatorial upwelling and polar subsidence.  The inter-hemispheric transport could be a circulation pattern analogous to Earth's Brewer-Dobson stratospheric circulation \citep[e.g.,][]{87andrews}, with air rising at low latitudes and descending at mid- to high-latitudes, inferred from the distribution of ozone and driven by planetary wave activity.  The single-cell idea for the Ice Giants should be compared to the better-studied gas giants, where the moderate-scale temperature contrasts observed in the tropospheres of Jupiter and Saturn are superimposed onto larger-scale hemispheric asymmetries (Fig. \ref{temp}), both non-seasonal on Jupiter \citep[e.g.,][]{18melin} and seasonal on Saturn \citep[see review by][]{18fletcher_book}. \citet{09guerlet} interpreted a local maximum in Saturn's hydrocarbons at $25^\circ$N as evidence for the descending branch of a Hadley-like circulation reaching into the stratosphere.  This was consistent with the seasonally-reversing circulation predicted by the model of \citet{12friedson}, with rising motion in the summer hemisphere and sinking motion in the winter hemisphere.  Low-latitude contrasts in hydrocarbons have also been observed on Jupiter \citep{10nixon, 16fletcher_texes, 18melin}, which could be related to a similar (albeit non-seasonal) Hadley circulation in the stratosphere.   But on Uranus and Neptune there's a problem with this large, single-cell idea.  

Such a pattern would create a cold equator (through adiabatic cooling) and warmer mid-latitudes, and would provide the windshear required to explain the results of \citet{18tollefson}\footnote{We note that the situation is more complex than this, as a cold equator and warm mid-latitudes are only needed if methane (and hence atmospheric density) are meridionally uniform.  However, the results of \citet{18tollefson} are somewhat degenerate, as the inclusion of an equator-to-pole depletion in methane by a factor of four allows for the warm equator and cool mid-latitudes that are actually observed, consistent with Neptune's ``upper cell'' of air rising at mid-latitudes and sinking over the equator.  This study makes it clear that both the temperature gradients and density gradients should be accounted for when trying to interpret vertical wind shear.}.  However, observations show that the equator is warm in the upper troposphere, requiring equatorial subsidence.  Furthermore, at mid-latitudes the upwelling implied by the cool upper tropospheric temperatures and sub-equilibrium para-H$_2$ are at odds with a finding from \citet{11karkoschka_ch4}, who required that downwelling must be occurring in the $45-50^\circ$S band to explain their Neptune methane depletion results.  This suggests that the methane hemispheric contrast is in fact a feature of the atmosphere below the 1-bar level, as indicated by the thick blue arrows in Fig. \ref{cartoon_icegiant}.  This also does not explain the seasonal re-appearance of Uranus' reflective bands near $38-58^\circ$ in both hemispheres, where upwelling of CH$_4$- and H$_2$S-laden air and subsequent condensation is a likely candidate \citep{14sromovsky}.  We suggest that these bands might be a feature of the atmosphere above the 1-bar level, and be related to the cool temperatures and sub-equilibrium para-H$_2$ there.  A multi-tier stack of circulation cells can be invoked in Fig. \ref{cartoon_icegiant} to reconcile these observations.

Just as the tropical upwelling of Jupiter and Saturn must extend deep (potentially deeper than the extratropical circulations), so too must the polar subsidence on the ice giants to explain the microwave brightness.  This is indicated by thick arrows penetrating all the way down through our altitude range in the polar regions.  However, \citet{14sromovsky} pointed out that polar subsidence would tend to inhibit the formation of condensate clouds there, counter to what is observed.  Small regions of upwelling, embedded within the generally subsiding region, have been proposed as an explanation, similar to the upwelling in the cyclonic belts of Jupiter.  \citet{14sromovsky} go one step further and suggest three stacked cells:  a deep cell producing the polar ammonia depletion, an intermediate cell producing upwelling H$_2$S condensates at high latitudes to produce the small puffy clouds, and then a top cell producing high-latitude methane depletion.  The question, however, is \textit{whether the puffy polar clouds can form in a region of general subsidence}, obviating the need for a region of convergence and upwelling at high latitudes.

Intriguingly, if the ice giant `tropics' are analogous to the cyclonic belts of Jupiter - warm and subsiding above the mid-plane, with upwelling below the mid-plane - maybe this would be a region where statically-stable layers associated with their water clouds would be thinner, promoting moist convection.  Might this be a location for Uranian and Neptunian lightning, as indicated in our schematic in Fig. \ref{cartoon_icegiant}?  No remote sensing observations can probe down to these watery depths.  This is in contrast to the mid-latitudes, where upwelling associated with methane or H$_2$S convection might explain the observable cloud features.

\section{Conclusions and Perspectives}
\label{conclusion}

The complexity of a multi-tier, stacked-cell structure for giant planet atmospheres, separating the mid-tropospheric weather layer (where eastward zonal jets are forced by eddy momentum flux convergence) from the stably stratified upper troposphere (where jets decay due to a poorly-understood eddy- or wave-induced dissipation), flies in the face of Occam's razor.  In this respect, our review is intended to be thought-provoking, and we have taken the liberty of generalisations in comparing the two gas giants and the two ice giants.  Multi-tier structures have emerged in the \textit{observational} literature for all giant planets over the past two decades, as a means of reconciling conflicting measurements of eddies, temperatures, and compositional tracers \citep{00ingersoll, 04ingersoll, 05showman, 09delgenio, 11fletcher_vims, 14sromovsky, 18showman, 18tollefson}.  Furthermore, the \textit{numerical simulation} literature have focused primarily on the generation of jets via eddy momentum flux convergence, and have produced patterns of upwelling and subsidence (and resulting temperature and tracer fields) that are more akin to our mid-tropospheric cell than the upper-tropospheric cell \citep[e.g.,][]{08lian, 09schneider, 10lian, 20spiga, 18young}.  Indeed, as the majority of evidence for upwelling and subsidence support the `classical view' of rising motions in zones and falling in belts, this gives the impression that sophisticated simulations are not reproducing reality \citep[although there are some analytical exceptions, e.g.,][]{09zuchowski}.

However, this conflict is in the eye of the beholder, as the interpretation of a remote-sensing dataset depends on \textit{where} in the atmosphere it is probing.  Deep-probing measurements (e.g., microwave, 5-$\mu$m) might sense the mid-tropospheric cell, shallow-probing measurements (e.g., reflected sunlight, mid-infrared) might sense the upper tropospheric cell.  The interface between these counter-rotating cells (also referred to as the `mid-plane') seems to be conveniently located near the top-most clouds where the eddy fluxes have been measured.  However, this seemingly contrived coincidence could be explained by the increasing importance of sunlight leading to the stable stratification of the upper troposphere.  We challenge future numerical simulations to attempt a reconciliation with the observations of the upper-tropospheric cell (as well as the zonal jet decay), as the observations in this region could provide helpful constraints on the assumptions underpinning the modelling.  Ultimately, a true test of our understanding of these belt/zone circulation patterns will emerge from our ability to simulate these planets with GCMs.

Progress is also required from observations, in particular those that probe the mid-troposphere.  The situation is perhaps best summed up by a quote from \citet{09delgenio}: ``...until remote sensing down to the water condensation level and below becomes a reality... it will be difficult for the numerous theories of [giant planet] circulation to be regarded as anything more than simply plausible ideas."  Juno's microwave observations could open up this prospect \citep{17li}, but need to be analysed in tandem with other wavelengths sensing temperature, clouds, and composition.  Cassini's near- and mid-infrared observations may probe the interface between Saturn's two cells, but need to be analysed simultaneously to determine whether phosphine, arsine, and ammonia are really sensing different meridional overturning patterns \citep{11fletcher_vims}.   Future missions to the ice giants must find a way to probe the circulation patterns below the top-most clouds of methane and H$_2$S ice \citep[e.g.,][]{18irwin_h2s}.  And continued monitoring of temporal variations and episodic outbursts in the belts and zones \citep{18sanchez_storm, 17fletcher_cycles, 18antunano_EZ, 19antunano} could reveal insights into the shifting balance between the meridional circulation cells, and the forces determining their quasi-periodic timescales.

The challenge to observers and modellers is therefore to prove or disprove this complex multi-tiered circulation system, and provide a better explanation for the multiple conflicting interpretations of giant planet observations.  We end this review back at the beginning, by noting that the multi-tiered circulation is not so different from the terrestrial case.  Earth's troposphere (including both the Hadley cell and the transport in the mid-latitudes by eddies) is thermally direct, with the temperature decreasing from the equator to the poles.  Considering thermal wind balance, this implies eastward winds that increase with altitude, in a ``lower cell''.  But these eastward jets do not increase throughout the stratosphere, instead they decay with altitude.  From thermal wind balance, this implies a temperature field that increases towards the poles in an annual mean sense.  This reversal in the sign of the temperature gradient in the ``upper cell'' occurs because of a thermally indirect circulation driven by waves that propagate from the troposphere into the lower stratosphere, where they break and are absorbed.  So, this picture of Earth's lower stratosphere is not so different from the ``classical'' picture of Jupiter's belts and zones above the clouds, namely a thermally indirect wave-driven circulation that produces temperature patterns which, in thermal-wind balance, cause the jets to decay with altitude.

\begin{acknowledgements}
Fletcher was supported by a Royal Society Research Fellowship and European Research Council Consolidator Grant (under the European Union's Horizon 2020 research and innovation programme, grant agreement No 723890) at the University of Leicester.  Kaspi was supported by the Israeli Space Agency.  The material in this review has benefited from conversations with Mark Hofstadter, Rick Cosentino, Roland Young, Cheng Li and Aymeric Spiga, and we are grateful to them for clarifying the interpretations of their GCM models.  We are extremely grateful to two anonymous reviewers for their careful and constructive criticisms.  We acknowledge funding from the \textit{Understanding the Diversity of Planetary Atmospheres} workshop at the International Space Science Institute (ISSI) in November 2018 for inspiring this review.

\end{acknowledgements}

\bibliographystyle{aps-nameyear}      
\bibliography{references.bib}                

%
%



\end{document}